\documentclass[10pt,prd,twocolumn,preprintnumbers,amsmath,amssymb]{revtex4}
\usepackage[latin1]{inputenc}
\usepackage[gen]{eurosym}
\usepackage{graphicx}
\usepackage{floatflt}
\usepackage{dcolumn}
\usepackage{bm}
\begin{document}
\title{Cosmology of fermionic dark matter}
\author{Tillmann Boeckel}
\affiliation{Institut für Theoretische Physik / Astrophysik, J.W. Goethe Universität\\ Max-von-Laue-Straße 1, D-60438 Frankfurt am Main, Germany}
\author{Jürgen Schaffner-Bielich}
\affiliation{Institut für Theoretische Physik / Astrophysik, J.W. Goethe Universität\\ Max-von-Laue-Straße 1, D-60438 Frankfurt am Main, Germany}
\date{\today}
\begin{abstract}
We explore a model for a fermionic dark matter particle family which decouples from the rest of the particles when at least all standard model particles are in equilibrium. We calculate the allowed ranges for mass and chemical potential to be compatible with big bang nucleosynthesis (BBN) calculations and WMAP-data for a flat universe with dark energy ($\Omega^0_{\Lambda}$ = 0.72, $\Omega^0_{M}$ = 0.27, $h$ = 0.7). Futhermore we estimate the free streaming length for fermions and antifermions to allow comparison to large scale structure data (LSS). We find that for dark matter decoupling when all standard model particles are present even the least restrictive combined BBN calculation and WMAP results allow us to constrain the initial dark matter chemical potential to a highest value of 6.3 times the dark matter temperature. In this case the resulting mass range is at most 1.8 eV $\leq$ m $\leq$ 53 eV, where the upper bound scales linearly with $g^s_{eff}(T_{Dec})$. From LSS we find that similar to ordinary warm dark matter models the particle mass has to be larger than $\sim$ 500 eV (meaning $g^s_{eff}(T_{Dec}) > 10^3$) to be compatible with observations of the Ly $\alpha$ forest at high redshift, but still the dark matter chemical potential over temperature ratio can exceed unity.	
\end{abstract}
\maketitle

\section{Introduction}
In recent years observations and the high precision analysis of the cosmic microwave background radiation \cite{Spergel07}, of large scale structure \cite{Tegmark06}, of the abundancies of light elements in the universe \cite{Steigman06} and of the evolution of the Hubble parameter by observation of the luminosity-redshift relation of supernovae of type Ia \cite{Astier05}, \cite{Riess06} have been strongly indicating that the universe is flat and mainly consists of two unknown components, dark matter and dark energy. Dark matter appears to be consisting of non-relativistic particles that only interact gravitationally and perhaps by weak interactions. In addition there has been a recent independent support for the dark matter paradigm by combined x-ray and weak lensing observation of the aftermath of a collision of two galaxies clusters \cite{Clowe06}.

The nature of the second unknown component, dark energy, is even less clear, it is mostly characterized by having a negative pressure that allows to explain the observed late time exponential expansion of the universe. In the simplest models dark energy is just a constant vacuum energy. Combined with a sizeble amount of cold dark matter this constitues the so called $\Lambda$CDM or concordance model.

For an alternative to dark energy using a large scale local inhomogeneity see e.g. \cite{Alnes05,Alnes06}. Some authors have even tried to unify dark matter and dark energy within so called Chaplygin gas models \cite{Amendola03,Sen05,Bertolami05,Gorini05}. This approach faces the difficulty of having a speed of sound increasing at late times which makes it probably unable to reproduce the matter power spectrum of density fluctuations on smaller scales.

Also there has been the phenomenological MOND approach by Milgrom \cite{Milgrom83} using a modified Newtonian force law and the more recent TeVeS theory by Bekenstein \cite{Bekenstein04} that tried to explain the observations without the need for dark matter. Still most of the work using these theories has been done on galactic rotation curves, stellar velocities in dwarf galaxies and the Pioneer anomaly. It is unclear whether they even succeed in explaining the observations at these scales so they still have to prove being an viable cosmological alternative to dark matter. For example Ref.~\cite{Slosar05} find that a simple MONDian model with a low matter density gives a rather bad fit to the power spectrum of the CMBR anisotropies when including the third acoustic peak extracted from BOOMERanG and WMAP first year data. A quite interesting recent approach to explain at least galactic rotation curves without the need for exotic dark matter or a new theory of gravitation has been done by Cooperstock et al. \cite{Cooperstock06} using a axially symmetric metric and linearized general relativity for disk galaxies. Although the mentioned different approaches may be able to remove the need for dark matter on and below galactic scales it is highly unclear whether they can do so for the CMB, large scale structure or weak lensing observations. All in all there is at present no convincing alternative to the dark matter paradigm that can explain all the mentioned independent observations.

The $\Lambda$CDM model is successful in explaining most of the observations although it currently predicts an excessive amount of low mass dark matter halos orbiting larger galaxies in both linear structure formation approaches as well as in numerical n-body simulations, as found by Ref.~\cite{Klypin99,Moore99}. Another problem of the CDM model is the prediction of cuspy inner density profiles of dark matter halos in dwarf and spiral galaxies, that was not confirmed by observations (see e.g.~Ref.~\cite{Gentile04,Salucci07,Gentile07}). Instead it was found that the core density of dark matter must to be rather constant. It is unclear whether these problems can be resolved in more realistic simulations/models that include for example more effects from the baryonic matter, or if these actually show that the dark matter properties are different from the standard CDM case. For example Ref.~\cite{Dave01} proposed self-interacting dark matter as a solution to the cuspy-core problem, but Ref.~\cite{Hennawi02} found that the interesting parameter range in which self-interacting dark matter could flatten halo cores can probably be excluded by other observations. Another solution could be warm dark matter (WDM) as studied by Ref.~\cite{Bode01}, where the core densities are reduced by the higher velocity dispersion of WDM.

It is usually assumed that the dark matter constituent, whether it is bosonic (a candidate would be the axion) or fermionic (for example the neutralino, the gravitino or a massive sterile neutrino), initially has a negligible chemical potential (or to be more exact that the condition $T \gg \mu - m$ is fullfilled). The reason for this assumption is the expectation that the matter-antimatter asymmetry in the universe is of the same magnitude for all particle species. So spin statistics are mostly ignored for dark matter and the usual approach for cold dark matter is then a free Boltzmann gas. In contrast to that we will use a free Fermi gas in our approach to find constraints on the allowed matter-antimatter-asymmetry through the related chemical potentials and the resulting mass window of a thermal fermionic dark matter relic from cosmological observations. We assume an additional conserved quantum number beyond those of the standard model to keep the particle stable, but this is common in many dark matter models. We will also explore possible implications for structure formation for such a particle species.

\section{Initial assumptions and motivation}
We assume a thermal distribution of a fermionic particle family with 4 degrees of freedom (Spin $\frac{1}{2}$) was created asymmetrically (i.e. a surplus of particles over antiparticles or the other way around) at an early stage somewhere around the GUT-scale (approx. $10^{15}$ GeV). The assymmetric process then became ineffective and the surplus of fermions was conserved leading to a non-zero chemical potential of these particles. We then assume thermal decoupling at a later stage while it was still ultra-relativistic. Thus the particles will be unaffected by the following numerous phase transitions and the subsequent reduction of the number of equilibrium particle species and their degrees of freedom $g_i$. Assuming overall entropy conservation from the dark matter decoupling scale up to today demands:
\begin{equation}
	\left(g^s_{eff} a^3 T^3\right)_{Dec} = \left(g^s_{eff} \:\: a^3 T^3\right)_{later}
\end{equation}
where
\begin{eqnarray}
	\label{14} g^s_{eff} = \sum_{i = bosons} g_i \left(\frac{T_i}{T}\right)^3 + \frac{7}{8} \sum_{i = fermions} g_i \left(\frac{T_i}{T}\right)^3
\end{eqnarray}
which allows to write the entropy using just one overall temperature (usually the photon temperature). In eq.~(\ref{14}) it is assumed that all the standard model particles (or whatever other particles may be present at dark matter decoupling) have negligible chemical potentials at least at the GUT scale and at the time of BBN. The temperature of equilibrium particles therefore is shifted upwards compared to the temperature of decoupled particles (like our dark matter particles) leading to the temperature relation:
\begin{equation}
	\label{6}T = T_{DM} \left(\frac{g^s_{eff}(T_{Dec})}{g^s_{eff}(T)}\right)^{1/3}
\end{equation}
The equilibrium relations $\mu = - \bar\mu$ and $T = \bar T$ will still be fullfilled as long as both fermions and antifermions are ultra-relativistic (where barred quantities refer to antiparticles), although they are decoupled and  no interactions take place anymore. The transition from ultra-relativistic to non-relativistic behaviour of both fluids needs to be treated numerically, meaning that we can start our numerical calculation at relatively late times (i.e. after BBN) and simply add the energy density, pressure etc.~of the underlying photon, neutrino and baryon fluids where their behaviour and magnitude is known from standard cosmology, WMAP-data and BBN-calculations. Neutrinos are treated as massless for simplicity.
   
\section{Analytic limits for energy density and equation of state}
\label{energy}
The energy density and equation of state (EoS) of a non-interacting Fermi gas with non-zero chemical potential can be calculated analytically in the ultra-relativistic as well as in the non-relativistic limit, an outline of the derivations will be given below, a more detailed description can be found in \cite{Sagert06}.\\
The energy density $\epsilon^+ = \epsilon + \bar \epsilon$ and pressure $p^+ = p + \bar p$ and net number density $n^- = n - ~ \bar{n}$ of a free non-interacting Fermi gas in chemical equilibrium are given in natural units ($c=\hbar=k_B=1$) by the following integrals over the momentum $\vec k$ of the particle:
\begin{eqnarray}
	\label{1} \epsilon^+ = \frac{g}{(2\pi)^3}\left( \int\limits_{0}^{\infty}d^3\vec{k} \: E(\vec{k}) \: \frac{1}{\exp{(\frac{E-\mu}{T}})+1}\right.\nonumber \\ \left. +\int\limits_{0}^{\infty}d^3\vec{k} \: E(\vec{k}) \: \frac{1}{\exp{(\frac{E- \bar\mu}{T}})+1} \right)
\end{eqnarray}
\begin{eqnarray}
	\label{13} p^+ = \frac{g}{(2\pi)^3}\left( \int\limits_{0}^{\infty}d^3\vec{k} \: \frac{k^2}{3 E(\vec{k})} \: \frac{1}{\exp{(\frac{E-\mu}{T}})+1}\right.\nonumber \\ \left. +\int\limits_{0}^{\infty}d^3\vec{k} \: \frac{k^2}{3 E(\vec{k})} \: \frac{1}{\exp{(\frac{E-\bar\mu}{T}})+1} \right)
\end{eqnarray}
\begin{eqnarray}
	\label{7} n^- = \frac{g}{(2\pi)^3}\left( \int\limits_{0}^{\infty}d^3\vec{k} \: \frac{1}{\exp{(\frac{E-\mu}{T}})+1} \right.\nonumber \\ \left.-\int\limits_{0}^{\infty}d^3\vec{k} \: \frac{1}{\exp{(\frac{E -\bar\mu}{T}})+1} \right)
\end{eqnarray}

Where the energy-momentum-dependence reads: $E(\vec{k}) = \sqrt{\vec{k}^2+m^2}$. Here $g$ denotes the degeneracy factor (i.e.~in our case $g=2$ for spin), not counting particles/antiparticles since this has already been accounted for by the two separate integrals.

\subsection{Ultra-relativistic limit}
\label{UR-limit}In the ultra-relativistic limit $E\approx k$ the integrals are solvable giving us the equation of state $p = \epsilon / 3$:
\begin{eqnarray}
	\epsilon^+ \approx \frac{g}{2\pi^2}\left( \int\limits_{0}^{\infty}dk \: \frac{k^3}{\exp{(\frac{k-\mu}{T}})+1}\right.\nonumber \\ \left. +\int\limits_{0}^{\infty}dk \: \frac{k^3}{\exp{(\frac{k-\bar\mu}{T}})+1} \right) = 3 p^+
\end{eqnarray}
by using the subtitutions
\begin{equation}
	\nonumber x_1 = \frac{k-\mu}{T},\:\: x_2 = \frac{k-\bar\mu}{T} 
\end{equation}
for the first and the second integral, respectively. By renaming $x_1$ and $x_2$ to $x$ and rearranging one ends up with the following three integrals (assuming $\mu = - \bar\mu$ to be true as said before):
\begin{eqnarray}
	\epsilon^+ \approx \frac{g \: T^4}{2\pi^2}\left( \int\limits_{0}^{\infty}dx \: \frac{2x^3}{\exp{x}+1}\right.\nonumber \\ \left. + \frac{\mu^2}{T^2} \int\limits_{0}^{\infty}dx \: \frac{6x}{\exp{x}+1} + \int\limits_{-\mu/T}^{0}dx \: \left(x + \frac{\mu}{T}\right)^3 \:  \right)\nonumber
\end{eqnarray}
So finally we arrive at the ultrarelativistic limit of the energy density of a free Fermi gas with non-zero chemical potentials :
\begin{equation}
	\label{3}\epsilon^+ \approx \frac{g \: 7 \pi^2}{120} \: T^4 + \frac{g}{4}\: \mu^2 \: T^2 + \frac{g}{8\pi^2}\:\mu^4 
\end{equation}
This formula can also be derived using the so-called Sommerfeld-expansion for eq.~(\ref{1}), which is a power series in $(T/\mu)^2$, as for $m = 0$ all terms above second order will vanish and one obtains eq.~(\ref{3}). 
\subsection{Non-relativistic limit}
\label{NR-limit}In the non-relativistic limit $E \approx m$ and consequently $\epsilon \approx m\:n$ (and $\bar \epsilon \approx m\:\bar n$). But there are still two limiting cases to distinguish. Both limits refer to particles as well as to antiparticles, since both fluids are still present in the non-relativistic regime. 

\subsubsection{Non-relativistic degenerate limit} 
In the degenerate case with $(\mu - m) \gg T$, only particles with a momentum higher or equal the Fermi momentum $k_F$ can be added to the system since all momentum states with lower energy have already been filled with particles. Eq.~(\ref{1}) simplifies then to:
\begin{eqnarray}
	\label{2}\epsilon \approx \frac{g}{2\pi^2} \int\limits_{0}^{k_F}k^2 dk \: m = \frac{g\: m\: k^3_F}{6 \pi^2} \\\nonumber k_F = \left(\frac{6 \pi^2 \epsilon}{g\: m}\right)^{1/3} = \sqrt{\mu^2 - m^2}
\end{eqnarray}
Using eq.~(\ref{2}) one sees that the pressure scales as
\begin{equation}
	p \approx \frac{g}{2\pi^2} \int\limits_{0}^{k_F}k^2 dk \: \frac{k^2}{3 m} = \frac{g \:k_F^5}{30 \pi^2 m} \propto  \epsilon^{5/3}
\end{equation}
which is a polytropic equation of state of the form 
\begin{eqnarray}
p \approx B \: \epsilon^{\gamma} \:\: \mbox{with}\:\: \gamma = \frac{5}{3}\:\: \mbox{and} \:\:	B = \left(\frac{36\: \pi^4}{125\: g^2\: m^8}\right)^{1/3}
\end{eqnarray}

\subsubsection{Non-relativistic non-degenerate limit}
In the non-degenerate case $(\mu - m) \ll T$, the Fermi-Dirac-distribution function can be well approximated by the Maxwell-Boltzmann-limit so that: 
\begin{eqnarray}
	 \epsilon \approx \frac{g}{2\pi^2} \int\limits_{0}^{\infty}dk k^2 \: m \: \exp{\left(-\frac{k^2}{2mT}\right)}\nonumber\\ \times \: \exp{\left(\frac{\mu - m}{T}\right)} = m n
\end{eqnarray}
Using the substitution $x = \frac{k^2}{2mT}$ one arrives at
\begin{equation}
 	\label{11}\epsilon \approx g\:m \left(\frac{mT}{2\pi}\right)^{3/2} \exp{\left(\frac{\mu - m}{T}\right)}= m n
\end{equation}
whereas a similar calculation for the pressure gives
\begin{equation}
 	p \approx g\:T \left(\frac{mT}{2\pi}\right)^{3/2} \exp{\left(\frac{\mu - m}{T}\right)} = T n
\end{equation}
Again, the equation of state is a polytrope of the form  
\begin{eqnarray}	 
	p \approx C \epsilon^{\gamma}\:\: \mbox{with}\:\:\gamma = \frac{5}{3}\:\: \mbox{and} \:\:	C = \left(\frac{ 8\pi^3}{g^2\: m^8 \: D^2}\right)^{1/3} 
\end{eqnarray}
Interestingly, this leads to almost the same equation of state because the exponential factor $D\equiv\exp{\left(\frac{\mu - m}{T}\right)}$ turns out to become approximatly constant in the non-relativistic limit since $\mu - m \propto a^{-2}$ and $T \propto a^{-2}$, which we will outline below in more detail.

\subsection{Generalized Chaplygin gas}
\label{Chaplygin gas}A polytrope resembles a generalized Chaplygin gas which has an equation of state of the form
\begin{equation}
	p = - \frac{A}{\epsilon^{\Gamma}}  
\end{equation}
where A and $\Gamma$ are normaly taken to be positive. Then the Chaplygin gas can be used as an unified model for dark matter and dark energy, see e.g. in \cite{Amendola03,Sen05,Bertolami05,Gorini05}. The scaling behaviour of the generalized Chaplygin gas can be found by solving
\begin{equation}
	\label{9}\frac{d\epsilon}{da} + \frac{3(\epsilon + p)}{a} = 0
\end{equation}
which can be deduced from energy-momentum conservation. The general solution is :
\begin{equation}
	\epsilon = \left(A + \frac{\epsilon_0^{(\Gamma + 1)} - A}{a^{3(\Gamma + 1)}}\right)^{1/(\Gamma + 1)}
\end{equation}
where $\epsilon_0$ is the energy density at $a = 1$. We will now examine the non-relativistic Fermi gas solution in comparison. For $\Gamma = -5/3$ the solution is
\begin{equation}
	\epsilon = \left(A + (\epsilon_0^{-2/3} - A)a^{2}\right)^{-3/2}
\end{equation}
For negative A the energy density runs into a pole at the scale parameter $a_{pole} = \sqrt{-A /(\epsilon_0^{-2/3} - A)}$. The energy density of a free Fermi gas should not become infinite in a finite volume. The way out of this problem is to consider that the gas will become relativistic, i.e. the EoS should change to that of an ultrarelativistic gas $p =\epsilon /3$, which can be estimated in the following way. First we formulate the constraint in terms of the scale parameter: can $a_{Rel}$ be smaller than $a_{Pole}$? The point where both limits of the non-relativistic and the ultra-relativistic EoS approach each other
\begin{equation}
	p_{Rel} \approx \frac{\epsilon_{Rel}}{3} \approx -A \epsilon^{5/3}_{Rel}
\end{equation}
leads to
\begin{equation}
	\epsilon_{Rel} = \left(-\frac{1}{3A}\right)^{3/2}
\end{equation}
for the energy density of the transition from the non-relativistic to the relativistic EoS. Now we can use the non-relativistic scaling to eliminate $\epsilon_{Rel}$ 
\begin{equation}
	\frac{\epsilon_{Rel}}{\epsilon_{0}} \approx	\left(\frac{a_{Rel}}{a_0}\right)^{-3}
\end{equation}
Normalizing to $a_0 = 1$ as before we arrive at the consistency condition $a_{pole} > a_{Rel}$ in the form
\begin{equation}
	\sqrt{\frac{-A}{\epsilon_0^{-2/3} - A}} ~ > ~ \left(\epsilon_0(-3A)^{3/2}\right)^{1/3}
\end{equation}
which can be rewritten to
\begin{equation}
	\epsilon_0^{2/3} A >  \frac{2}{3}
\end{equation}
This condition is never fullfilled since $A$ is negative, meaning that the gas will become relativistic before it can run into the pole. 
  
\section{Chemical potential and temperature}

\subsection{Ultra-relativistic behaviour}
After chemical decoupling of the dark matter particle/antiparticle distribution from the standard model particle background, the net number, i.e. the surplus of particles over antiparticles of the fermion distribution in a comoving volume, will be frozen in (as well as the number of particles and antiparticles separately). This can be expressed by the conservation equation
\begin{equation}
	\label{8}\dot{n}^- + 3 H n^- = 0 
\end{equation}
where $H$ is the Hubble parameter, leading to a simple dependence on the scale parameter a
\begin{equation}
	\label{4} n^-\propto a^{-3}
\end{equation}
The ultrarelativistic limit of the net number density can be calculated as a derivative of the pressure or the energy density, using eq.~(\ref{3}):
\begin{equation}
	n^- = \frac{\partial p}{\partial \mu} = \frac{1}{3} \frac{\partial \epsilon}{\partial \mu} = \frac{g}{6}\: \mu \: T^2 + \frac{g}{6\pi^2}\:\mu^3 
\end{equation}
showing that the net number density goes to zero for a vanishing chemical potential as expected. Furthermore since the net number density has to scale like $a^{-3}$ for any value of T and $\mu$ we can deduce the ultrarelativistic scaling of the temperature and the chemical potential: 
\begin{equation}
	\label{10}T \propto a^{-1} \:\:\:\: \mbox{and} \:\:\:\: \mu \propto a^{-1}
\end{equation}
so that
\begin{equation}
	\epsilon \propto a^{-4} \:\:\:\: \mbox{and} \:\:\:\: p \propto a^{-4}
\end{equation}
like for standard relativistic particles (with $ \mu = 0$). This is of cause only true as long as the chemical potentials and temperatures of fermions and antifermions are well above the mass of the particle. For a Fermi gas in chemical equilibrium with a surplus of particles that becomes non-relativistic, particles and antiparticles annihilate until all antiparticles are gone (exponentially suppressed with respect to particles, since $\bar\mu \rightarrow - m$) while conserving the net number per comoving volume $a^3 \cdot n^-$. In our case not only the net number, but also the numbers of particles and antiparticles are conserved separately, forcing both chemical potentials to approach the mass and the temperatures of the non-relativistic limit to be different (see below for more details).
 
\subsection{Non-relativistic behaviour}
As soon as the temperatures of the fermion fluids approach the rest mass energy the scaling of the temperatures and chemical potentials will start to be different. In the following equations $n, T$ and $\mu$ refer to particles or antiparticles, not both. From eqs. (\ref{11}) and (\ref{4}) we see that in the non-degenerate limit
\begin{equation}
	n \approx C_1 \: T^{3/2} \exp{\left(\frac{\mu - m}{T}\right)} = \frac{n_0}{a^3}
\end{equation}
If we now assume that $\mu - m$ and $T$ just scale like powers of the scale parameter 
\begin{equation}
	\mu - m = C_2 / a^x \mbox{   and   } T = T_0 / a^y \\
\end{equation} 
where $C_1$, $C_2$, $T_0$, $n_0$, $x$ and $y$ are constants, we arrive at
\begin{equation}
	\frac{n_0}{a^3} = C_1 \: T_0^{3/2} a^{-3y/2} \exp{\left(\frac{C_2}{T_0}\: a^{y-x}\right)}
\end{equation}
or
\begin{equation}
	a^{x-y} \left( \ln{\frac{n_0}{C_1 \: T_0^{3/2}}} + \frac{3}{2}(y-2)\ln{a} \right) =  \frac{C_2}{T_0}
\end{equation}
This can only be fullfilled for arbitrary scales if $x = y = 2$, meaning that
\begin{equation}
	\label{12}T \propto a^{-2} \:\:\:\: \mbox{and} \:\:\:\: (\mu - m) \propto a^{-2}
\end{equation}
Note that this infers that the Boltzmann factor $\exp{((\mu - m)/T)}$ stays constant in the non-relativistic limit. The energy density and pressure will then scale as $\epsilon \approx m n \propto a^{-3}$ and $p \approx B \epsilon^{5/3} \propto a^{-5}$ in the non-relativistic degenerate limit. We stress that this is exactly the same scaling as that of a Boltzmann gas, where $p \approx n T \propto a^{-5}$ as $T \propto a^{-2}$. It is noteworthy that the evolution of energy density and pressure with the scale parameter does not change in both the ultrarelativistic and the non-relativistic limit when a finite chemical potential is added.

\subsection{Big bang nucleosynthesis}
\label{BBN}The most important constaint on a chemical potential for dark matter can be deduced from big bang nucleo\-synthesis (BBN) calculations. The abundances of light elements are sensitive to the expansion rate of the universe during BBN. The expansion rate on the other hand is determined by the energy density of relativistic particles (radiation domination) during this short epoch (a few MeV $ \geq T_{photon} \geq $ 30 keV) which is usually expressed in terms of the effective number of neutrino families $N_{\nu}$, for a detailed description see Ref.~\cite{Steigman06}. For standard cosmology $N_{\nu} = 3$ (or to be more precise $N_{\nu} = 3.046$, which is due to the small overlap of neutrino decoupling and $e^+e^-$- annihilation) while the values of $N_{\nu}$ deduced from comparison of the observed abundances of light elements to BBN-calculations and from WMAP data differ and have relatively large error bars. For example the authors of Ref.~\cite{Cyburt05} find $N_{\nu} = 3.08 ^{+1.55}_{-1.28}$ at $2\sigma$ error using a combined $^4$He-abundance and CMB results for the baryon abundance. On the other hand Ref.~\cite{Mangano07} arrives at $N_{\nu} = 5.2 ^{+2.7}_{-2.2}$ at $2\sigma$ error using CMB and large scale structure data. Ref.~\cite{Barger03} find a similar upper limit for $N_{\nu} = 3.2 ^{+4.8}_{-2.2}$ at $2\sigma$ error using WMAP and BBN including $^4$He- and deuterium-abundances. In a more recent paper de Bernardis et al.~\cite{Bernardis07} find $N_{\nu} = 3.7 ^{+1.1}_{-1.2}$ at $2\sigma$ using CMB data, LSS, Supernova data and other independent measurements of the Hubble parameter, which is more restictive than Ref.~\cite{Mangano07,Barger03}. So although $N_{\nu}$ is in any case compatible with the standard value of $N_{\nu} = 3$ it can be stressed that some of the observations seem to favour a larger value, as for example found in Ref.~\cite{Mangano07} by including large scale structure data. We will give upper and lower bounds on the chemical potential and particle mass to allow comparison to these different results. The standalone BBN results from Cyburt et al.~\cite{Cyburt05} are nevertheless the most relevant (and most restrictive) for our approach since all the other measurements of $N_{\nu}$ refer to later stages in the evolution of the universe at which our dark matter particle will already be non-relativistic and thus does not contribute to the effective number of relativistic particle species.\\ 
Next we will estimate the allowed window for the value of the chemical potential of the dark matter particle. One has to be careful with the definition of $N_{\nu}$ at this point since the important quantity is the temperature of the neutrino distribution, which will differ from the photon temperature at BBN. This is due to the annihilation of $e^+$ and $e^-$ below 0.5 MeV, which occurs after the decoupling of neutrinos at around 1 MeV, and the subsequent release of entropy almost completely into the photon fluid. So we find the following constraint on the energy density of a fermionic dark matter particle $\Delta \epsilon_{rel} \geq \epsilon_{DM}$:
\begin{eqnarray}
	\Delta N_\nu \frac{7 \pi^2}{120} T^4_{\nu} \geq g_{DM} \cdot \left(\frac{7 \pi^2}{120} T^4_{DM}\right.\nonumber \\ \left. + \frac{1}{4} T^2_{DM}\mu^2_{DM} + \frac{1}{8\pi^2} \mu^4_{DM}\right)\label{5}
\end{eqnarray}
\noindent where T$_{\nu}$ is the neutrino temperature. Using eq.~(\ref{6}) and defining

\begin{eqnarray}
	\Theta \equiv \left(\frac{g^s_{eff}(T_{Dec})}{g^s_{eff}(T \approx 1\mbox{MeV})}\right)^{4/3}
\end{eqnarray}
and consequently
\begin{eqnarray}
	\nonumber  \:\: \Theta_{SM}  = \left(\frac{106.75}{2+\frac{7}{8}\left( 2 \cdot 2 + 2 \cdot 3 \right)}\right)^{4/3} \approx 21.3
\end{eqnarray}
for decoupling when all standard model particles are present and in equilibrium. We arrive at the following constraint on $\mu_{DM}/T_{DM}$:
\begin{eqnarray}
	 \frac{\Delta N_\nu \Theta}{g_{DM}} \geq 1 + \frac{30}{7 \pi^2} \left(\frac{\mu_{DM}}{T_{DM}}\right)^2 +\frac{15}{7 \pi^4} \left(\frac{\mu_{DM}}{T_{DM}}\right)^4
\end{eqnarray}
Solving for the largest value of $\mu_{DM}/T_{DM}$ that still fullfills this relation one arives at
\begin{equation}
	\label{55}\left.\frac{\mu_{DM}}{T_{DM}}\right|_{max} = \sqrt{-\pi^2+\sqrt{\frac{7\pi^4}{15}\left(\frac{\Delta N_\nu \Theta}{g_{DM}}+\frac{8}{7}\right)}}
\end{equation}
This constraint is of course only valid for fermions that are still relativistic at BBN. It does not apply for particles that were non-relativistic at decoupling or non-thermally produced. This is due to the fact that the contribution of non-relativistic particles to the total energy density at BBN must be negligible not to overclose the universe today, consequently one can hardly probe $\mu_{DM}/T_{DM}$ for such particles via BBN.\\
\begin{figure}[htbp]
	\centering
		\includegraphics[width=8.6 cm]{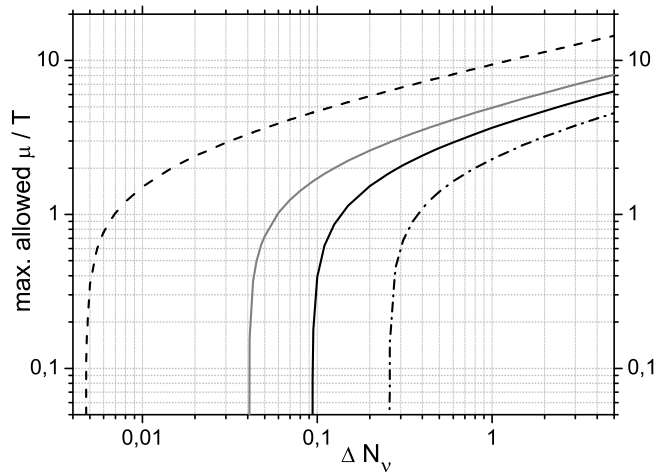}
	\caption{Maximum allowed dark matter chemical potential over temperature as a function of the number of additional eff\-ective neutrino families for different values of the effective number of degrees of freedom $g^s_{eff}$ at decoupling. The different lines correspond to the cases $g^s_{eff}(T_{Dec})$ = 1000 (dashed), 200 (solid gray), 106.75 (solid) and 50 (dash dotted)}
	\label{fig:Graph1}
\end{figure}
The constraints on the maximum allowed values for $\mu_{DM}/T_{DM}$ are plotted vs.$~\Delta N_{\nu}$ in figure \ref{fig:Graph1} ($g_{DM}$ is set to 2). For $\Delta N_{\nu} = 0.094$ a non-zero chemical potential can be excluded, for $\Delta N_{\nu} < 0.094$ even a non-degenerate fermion family can be excluded (the $T^4_{DM}$-term in eq.~(\ref{5}) always remains!). This limit will of course be shifted to a higher value of $\Delta N_{\nu}$ if $g^s_{eff}$ is smaller at decoupling than our reference value of 106.75. For $\Delta N^{max}_{\nu} = 5$ \cite{Mangano07, Barger03} the maximum allowed value for the chemical potential is 6.3 times the dark matter temperature. Futhermore even for $\Delta N_{\nu}$ = 1.63  \cite{Cyburt05}, the ratio $\mu_{DM}/T_{DM}$ can be as large as 4.4.
Figure \ref{fig:Graph1} also shows the allowed chemical potential for $g^s_{eff} = 50$ at DM-decoupling and for hypothetical values of 200 and 1000 showing that even such an enormous íncrease in $g^s_{eff}$ will allow a maximum chemical potential of only a factor of about 2 larger compared to decoupling while $g^s_{eff} = 106.75$.

In principle $g^s_{eff}(T_{Dec})$ could be much higher than the standard model value for example due to numerous additional super symmetric particles. E.g.~in a gas of strings in the very early universe there are exponentially many states present (see for example Ref.~\cite{Brandenberger07,Brandenberger06}). As we will see later very large values of $g^s_{eff}(T_{Dec})$ are essential for this approach to work. 

Of course there are other mechanisms that could lead to a variation in the extracted effective number of neutrino families from the standard value of $N_{\nu} = 3.046$. For example an evolving gravitational constant G could also lead to a difference in the expansion rate at BBN (see e.g.~Ref.~\cite{Kneller03}), as it could also be the case in cosmological models with more than the standard 4 spacetime dimensions (see e.g.~Ref.~\cite{Randall99}).   

\section{Numerical results}
In the following we examine one specific parameter set of a flat universe ($\Omega_{tot} = 1$) in our numerical treatment with the present day values $\Omega^0_{\Lambda} = 0.72$ (simple vacuum energy), $\Omega^0_{DM} = 0.236$, $\Omega^0_B = 0.044$ and $\Omega^0_{\gamma+\nu} = 8.5 \cdot 10^{-5}$, similar to the parameters obtained from WMAP-data using the $\Lambda$CDM-model \cite{Spergel07}. $\Omega^0_{X}$ as usual denotes the fraction of the critical energy density of the respective energy component today. We use a Hubble parameter of the value $H_0 = 70$ km s$^{-1}$ Mpc$^{-1}$.
We start our calculation of the Fermi-integral-sets  $(\epsilon, p, n)$ and $(\bar \epsilon, \bar p, \bar n)$ from eqs.~(\ref{1}), (\ref{13}), (\ref{7}) when the particles and antiparticles are still ultra-relativistic and use number and energy-momentum-conservation, eq.~(\ref{9}), for both fermions and antifermions separately to find the scaling of the chemical potentials and temperatures. When the latter have reached non-relativistic scaling behaviour with sufficient precision we continue using the non-relativistic expression of eq.~(\ref{11}).\\
We add photons, three massless non-degenerate neutrino families and baryons as particle background. To find the time dependence of the scale parameter $a$ we use the first Friedmann-equation for a flat universe (i.e.~with zero \mbox{ curvature):}
\begin{equation}
	\label{16}H^2 = \left(\frac{\dot a}{a}\right)^2 = \frac{8 \pi G}{3}\: \epsilon_{tot}
\end{equation}
Finally we calculate the free streaming length for fermions and anti-fermions to find additonal constraints on the particle mass and show that the scaling of the Jeans length and Jeans mass is exactly that of ordinary warm dark matter (WDM). 
 
\subsection{Mass range} 
\begin{figure}[hb]
	\centering
		\includegraphics[width=8.6 cm]{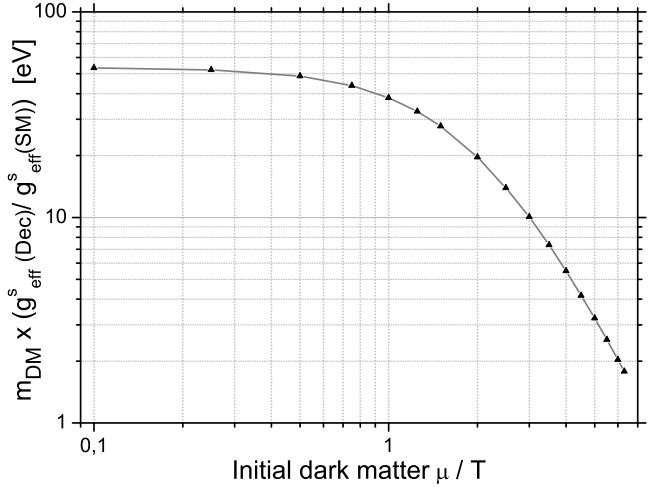}
	\caption{particle mass sufficient to close the universe for the allowed range of the initial chemical potential over temperature ratio}
	\label{fig:Graph2}
\end{figure}

\begin{figure}[ht]
	\centering
		\includegraphics[width=8.6 cm]{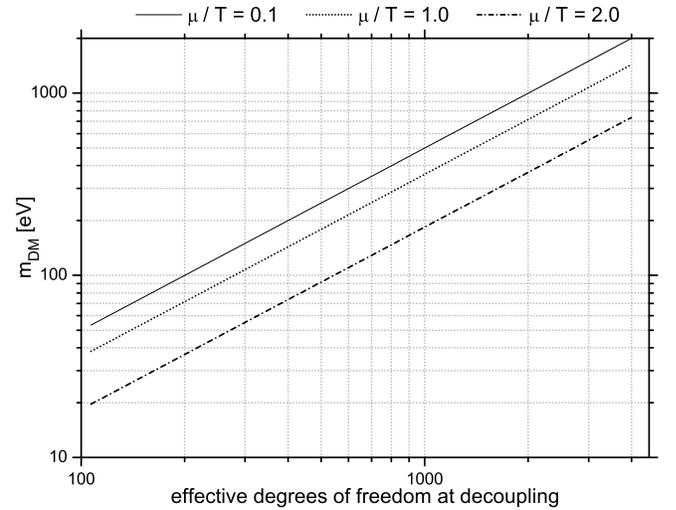}
	\caption{dark matter particle mass sufficient to close the universe vs. the degrees of freedom at decoupling for several values of the initial chemical potential over temperature ratio}
	\label{fig:Graph3}
\end{figure}

For a given initial value of $\mu_{DM} / T_{DM}$ we then use the condition $\Omega^0_{DM} = 0.236$ to pin down the particle mass which is sufficient to close the universe. Figure \ref{fig:Graph2} shows the allowed range of the mass, assuming $\Delta N^{max}_{\nu}$ = 5 (i.e $\mu_{DM} / T_{DM} \leq 6.3$), were we find 1.8 eV $\leq m \leq$ 53 eV. For $\Delta N^{max}_{\nu}$ = 1.63 (or a maximum initial value of $\mu_{DM} / T_{DM}$ = 4.38) the corresponding mass constraint is 4.4 eV $\leq m \leq$ 53 eV.
For small initial values of $\mu_{DM} / T_{DM}$ one finds the expected $T^4$-plateau due to the first term in the ultra-relativistic energy density, see eq.~(\ref{3}), which is independent of the chemical potential. The critical mass is determined just by the relativistic degrees of freedom at decoupling and the expected density today (or at any other arbitrary point in the non-relativistic regime). A good estimate of the upper limit on the mass (meaning for $\mu_{DM} / T_{DM} \approx 0$) is
\begin{eqnarray}
	\label{52}m_{max} \approx~ 3 \cdot T^0_{\gamma} \left(\frac{T_{DM}}{T_{\gamma}}\frac{\epsilon_{\gamma+\nu}}{\epsilon_{DM}}\right)_{T_{DM} > m}~  \frac{\Omega^0_{DM}}{\Omega^0_{\gamma+\nu}}~ h^2~  \nonumber\\ \approx~ 51 \left(\frac{4}{g_{DM}}\right)\left(\frac{g^s_{eff}(T_{Dec})}{106.75}\right) \left(\frac{\Omega^0_{DM}}{0.236}\right) \left(\frac{h}{0.70}\right)^2~ \mbox{eV}\nonumber
\end{eqnarray}
$g_{DM}$ is in this case equal to four since particles and antiparticles contribute equally to the energy density. The lower limit on the mass (in other words $\Delta N_{\nu} = \Delta N^{max}_{\nu}$) can be estimated by:
\begin{eqnarray}
	m_{min} \approx~ 3 \cdot T^0_{\gamma} \left(\frac{T_{DM}}{T_{\gamma}}\frac{\epsilon_{\gamma+\nu}}{\epsilon_{DM}}\right)_{T_{DM} > m} ~  \frac{\Omega^0_{DM}}{\Omega^0_{\gamma+\nu}}~ h^2~\nonumber\\\approx~1.6 \left(\frac{106.75}{g^s_{eff}(T_{Dec})}\right)^{1/3} \left(\frac{5}{\Delta N_{Max}}\right) \left(\frac{g_{DM}}{2}\right)\nonumber\\\times \left(\frac{\Omega^0_{DM}}{0.236}\right) \left(\frac{h}{0.70}\right)^2~ \mbox{eV}\nonumber
\end{eqnarray}
where in both cases we assume that the fermions become non-relativistic at $T \approx m / 3$. Most of the parameter range is dominated by the contribution of the second term in eq.~(\ref{3}), only for the values above $\mu_{DM} / T_{DM} \approx 4.4$ the third term $\propto \mu^4$ becomes dominant. For a larger effective number of degrees of freedom at decoupling the mass window is shifted to higher masses by a factor of $g^s_{eff}(T_{Dec})/ g^s_{eff}(SM)$, where $g^s_{eff}(SM)$ corresponds to all standard model particles. In this case the highest allowed value of the initial chemical potential will be larger (see section \ref{BBN}) meaning that the size of the allowed mass window will grow. It is also interesting to notice that the lower limit on the mass only has a slight dependence on the degrees of freedom at decoupling and is therefore always close to 1eV for all discussed values of $g^s_{eff}(T_{Dec})$.\\
In figure \ref{fig:Graph3} we show the dependence of the particle mass on the degrees of freedom at decoupling for initial values of $\mu / T = 0.1, 1.0, 2.0$, which is the most interesting range of parameters as we shall see later when discussing structure formation.

\subsection{Scaling of the temperature and chemical potential}
\begin{figure}[ht]
	\centering
		\includegraphics[width=8.6 cm]{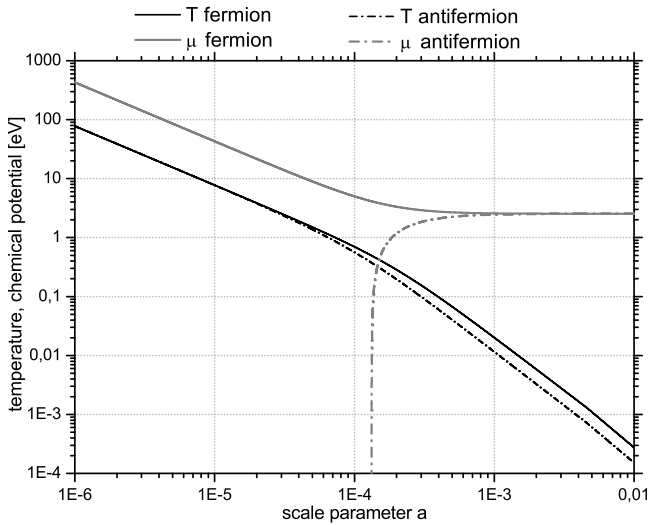}
	\caption{Temperatures and chemical potentials for an initial $\mu / T = 5.5$ and a mass of 2.54 eV}
	\label{fig:Graph4}
\end{figure}

\begin{figure}[hb]
	\centering
		\includegraphics[width=8.6 cm]{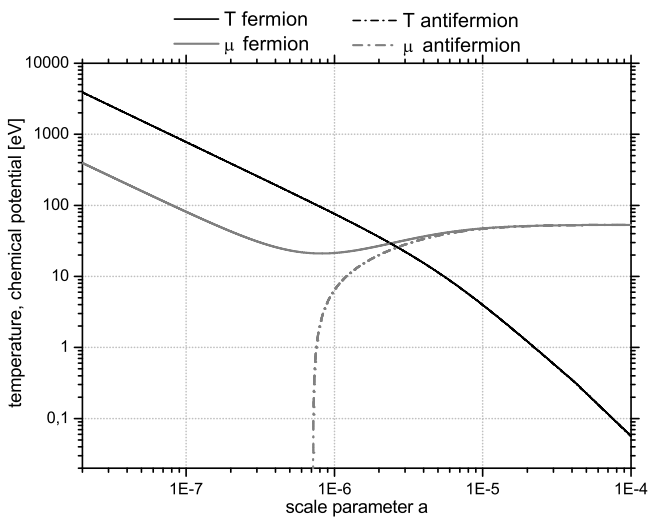}
			\caption{Temperatures and chemical potentials for an initial $\mu / T = 0.1$ and a mass of 53.3 eV}
	\label{fig:Graph5}
\end{figure}

Figures \ref{fig:Graph4} and \ref{fig:Graph5} show the scaling of the dark matter temperature and chemical potential for initial values of $\mu_{DM}/T_{DM} = 5.5, 0.1$ and a fermion mass of 2.54 eV and 53.3 eV, respectively. In the ultra-relativistic regime (i.e. for small values of $a$) we recover the simple $1/a$ - scaling for $\mu_{DM}$ and $T_{DM}$ as expected from eq.~(\ref{10}). In the non-relativistic regime we find that both chemical potentials are aproaching the mass and the temperatures scale as $ a^{-2}$ like in eq.~(\ref{11}).
Furthermore, one can see that the temperature of fermions will be higher than that of antifermions in the non-relativistic regime, which is mandatory for number conservation of both species (in figure \ref{fig:Graph5} the difference is too small to be visible).
The chemical potential of the antifermions, $\bar\mu$, approaches the mass from negative values with a similar rate as $\mu$ making it look like it is shooting up exponentially in the log-log-plot.\\
In figure \ref{fig:Graph5}, the case for small initial $\mu_{DM}/T_{DM}$, we see that both the particle and the antiparticle chemical potential are temporarily below the mass and change their scaling only when the temperature approaches the mass.
  
\subsection{Equation of state}
\label{EoS} As outlined in sections \ref{UR-limit} and \ref{NR-limit} one expects a radiation-like EoS in the ultra-relativistic limit as well as a polytropic equation of state with an exponent of 5/3 in the non-relativistic limit for both a degenerate and a non-degenerate Fermi gas. Actually one does not expect to see the non-relativistic degenerate limit since the highest allowed initial value of $\mu_{DM} / T_{DM}$ of 6.3 is only slightly higher than the value at which the $\mu^4$-Term in (\ref{3}) becomes larger than the $\mu^2T^2$-Term at a value of 4.4. So we actually examine only the medium- to non-degenerate region of a free Fermi gas.
\begin{figure}[ht]
	\centering
		\includegraphics[width=8.6 cm]{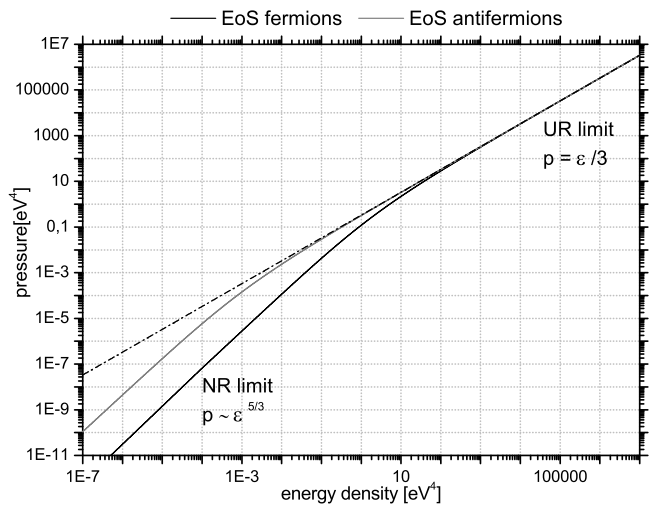}
	\caption{EoS for an initial $\mu / T = 5.5$ and a mass of 2.54 eV}
	\label{fig:Graph6}
\end{figure}
In Figure \ref{fig:Graph6} the EoS of fermions and antifermions for an initial $\mu / T = 5.5$ and a mass of 2.54 eV is shown. The EoS differs in the non-relativistic regime for a non-zero initial $\mu_{DM} / T_{DM}$ because the fermions are (at most) semi-degenerate while the antifermions are non-degenerate and very dilute (in comparison to the fermions). When comparing the fermion-pressure and the antifermion-pressure in the non-reltivistic regime at the same energy density for each particle-distribution the antifermions have a higher pressure because their temperature is higher for the same value of the energy density. The pressure of fermions is of course always larger than the pressure of antifermions at the same scale parameter for positive initial fermion chemical potential, because the energy density of antifermions can in fact be orders of magnitude smaller.
\subsection{Scaling of energy densities, number densities and pressures}
For a flat universe the total density is equal to the critical density at all values of the scale parameter $\epsilon_{tot}(a) = \epsilon_{crit}(a)$. This can be used to deduce the scaling of the individual normalized energy densities (usually denoted by $\Omega$). For example the normalized radiation energy density is given by
\begin{equation}
	\Omega_{\gamma+\nu} = \frac{\epsilon_{\gamma+\nu}}{\epsilon_{\gamma+\nu}+\epsilon +\bar\epsilon +\epsilon_{B} + \epsilon_{\Lambda}}
\end{equation}
where $\epsilon$ is the energy density of fermions and $\bar\epsilon$ is the one of antifermions.
\begin{figure}[ht]
	\centering
		\includegraphics[width=8.6 cm]{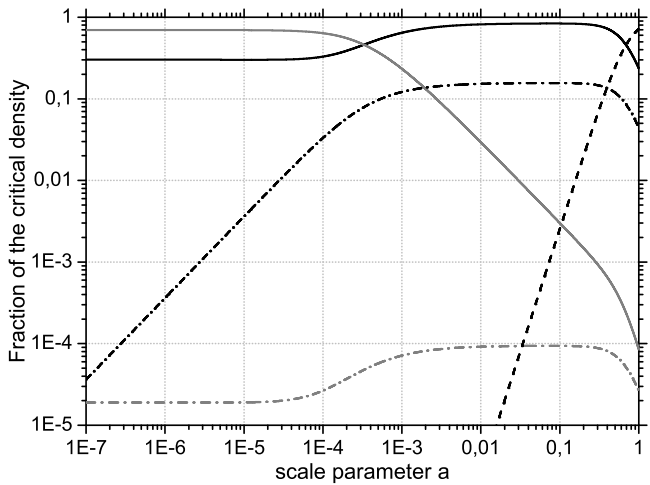}
	\caption{Fractions of the critical density for an initial $\mu / T = 5.5$, photons and neutrinos (solid gray), dark matter(fermions: solid, antifermions: dash dotted gray), baryons (dash dotted) and dark energy (dashed)}
	\label{fig:Graph7}
\end{figure}
\begin{figure}[ht]
	\centering
		\includegraphics[width=8.6 cm]{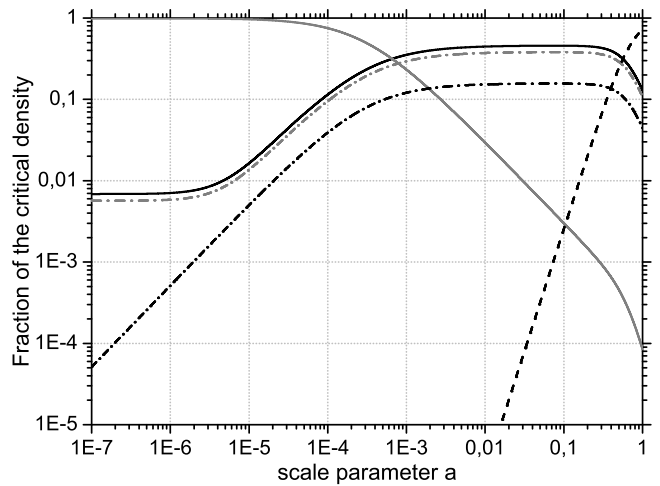}
	\caption{As fig.~\ref{fig:Graph7} but for an initial $\mu / T = 0.1$}
	\label{fig:Graph8}
\end{figure}
We show the scaling of the mean energy densities normalized to the critical density in figures \ref{fig:Graph7} and \ref{fig:Graph8} for the cases $\mu_{DM} / T_{DM}$ = 0.1 and 5.5. The plots show nicely what is often called the "coincidence problem", namely that dark energy seems to to become dominant just below a redshift $z = \frac{1}{a} - 1 \approx 0.5$. On can see that the contribution from fermions and antifermions is constant during radiation domination until they become non-relativistic and can constitute a substantial fraction of the total energy budget in the very early universe. Then their normalized energy densities grow linearly with $a$ until non-relativistic particles dominate. In contrast to that cold dark matter would in these plots scale just like the baryon density and exactly coincide with the scaling behaviour of the sum of the fermions and antifermion contributions in the matter dominated regime, i.e. the CDM contribution is highly suppressed at early times.\\In Figure \ref{fig:Graph9} we show an example of the scaling of energy density, pressure and number density of fermions and antifermions an initial $\mu_{DM} / T_{DM} = 5.5$. In the ultra-relativistic limit we recover  radiation like scaling of energy density and pressure being proportional to $a^{-4}$, in the non relativistic limit the energy densities scale like matter ($\epsilon \propto a^{-3}$) and the pressure drops rapidly ($p \propto a^{-5}$). The number densities always have a slope of $-3$ as demanded by number conservation of particles and antiparticles.
\begin{figure}[ht]
	\centering
		\includegraphics[width=8.6 cm]{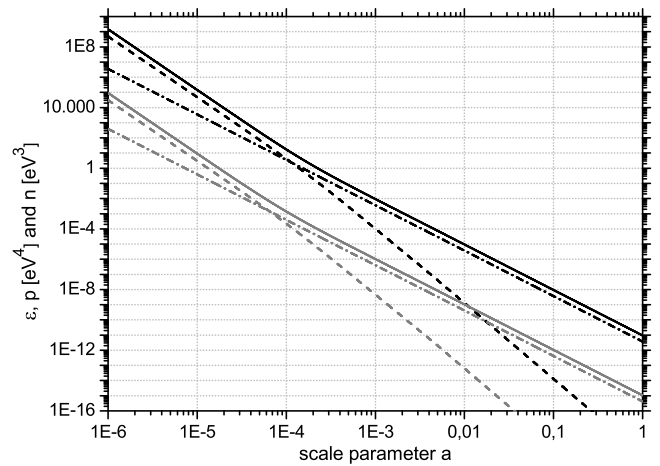}
	\caption{Evolution of the energy density (solid), pressure (dashed) and number density (dash dotted) of fermions (black) and antifermions (gray) for an initial $\mu / T = 5.5$ and a mass of 2.54 eV}
	\label{fig:Graph9}
\end{figure}

\subsection{Age of the universe}
It is generic of the cases studied ($\Omega_{\Lambda}^0$ = 0.72, $\Omega_{DM}^0$ = 0.236) to produce an almost fixed age of the universe, independent of the initial chemical potential over temperature ratio of the dark matter component. This is due to the fact that the dark matter has to become non-relativistic before the decoupling of photons and baryons occurs at around 380 000 years after the big bang. So any increase in the energy density at an earlier stage will only reduce the age of the universe by less than 380 000 years due to an increased Hubble parameter at earlier times.\\
So for $\Omega^0_{\Lambda} = 0.72$, $\Omega^0_{DM} = 0.236$ we find the age of the universe is the well known 13.7 Gyrs (as in $\Lambda$CDM), for example in a flat $\Omega^0_{M} = 1$ universe the age would be only 9.3 Gyrs(!). The latter case (although already being excluded by WMAP, if the spectrum of initial fluctuations is a simple power law \cite{Spergel07}) has the common problem of all flat matter dominated models to be in violation of some independent lower limits on the age of the universe. This is due to the today well messured value of the hubble parameter of around 70 km s$^{-1}$ Mpc$^{-1}$. For example from stellar evolution \cite{Chaboyer02} one finds a lower limit of 10 Gyrs on the age of $\omega$ Cen and Ref.~\cite{Paust07} finds M92 to be at least 13 Gyrs old. From the cooling of white dwarfs in globular clusters \cite{Hansen04} derives an age of more than 10.3 Gyrs (even at a 2$\sigma$ lower limit) on the age of M4, while in \cite{Hansen07} NGC 6397 is found to be 11.47$\pm$0.47 Gyrs.\\ A significantly lower value of the hubble parameter ($h < 0.5$) would be needed to solve this age problem and reach an age of more than 13 Gyrs for a flat matter dominaded universe also.

\subsection{Structure formation}
Now we will discuss more restrictive limits on the particle mass and degrees of freedom at decoupling which can be deduced from the shape of the matter power spectrum. In the second part of this section we will show that the general scaling of the relevant quantities for linear structure formation do not change when adding a finite chemical potential.
\subsubsection{Free streaming}
For collisionless particles (that have decoupled while being ultra relativistic) the most important damping scale is the free streaming length, as found by \cite{Boehm05}. The comoving free streaming length is given by 
\begin{equation}
	\label{17}\lambda_{FS} = \int_0^t{\frac{\left\langle v(t')\right\rangle}{a(t')}dt'} ~~~~~(\mbox{assuming  } t_{Dec} \ll t)
\end{equation}
which is basically the distance a particle can travel in an expanding background since decoupling. Interestingly Bond and Szalay \cite{Bond83} found that the relevant comoving damping scale from free streaming of collisionless particles can be estimated by 
\begin{equation}
	\label{18}\lambda_{FS} = 3.2 \left.\frac{\left\langle v(t)\right\rangle \: t}{a(t)}\right|_{Max}
\end{equation}
which is normaly in between the scale where particles become non-relativistic and the scale where matter starts to dominate.
The mean velocity of fermions can be calculated from the distributions function
\begin{equation}
	\left\langle v\right\rangle = \frac{g_{DM}}{(2 \pi)^3 n}\int_0^{\infty}{d^3\vec{k} \: \frac{k}{E(\vec{k})} \: \frac{1}{\exp{(\frac{E-\mu}{T}})+1}}
\end{equation}
The corresponding mass enclosed in a sphere of radius $\lambda_{FS}$ is the free streaming mass:
\begin{equation}
	M_{FS} = \frac{4\pi}{3}\lambda_{FS}^3 \bar\rho_M
\end{equation}
From the comparison of numerical n-body simulations of warm dark matter and cold dark matter Ref.~\cite{Narayanan00} claim that a (thermal) warm dark matter particle of a mass below 750 eV would be incompatible with observations of the Ly $\alpha$ forest at a redshift of $z \approx 3$. On the other hand, Boehm et al.~\cite{Boehm05A} find with higher resolution simulations that warm dark matter particle masses down to 600 eV would produce a power spectrum that is virtually indistinguishable from a simple cold dark matter spectrum below $z \approx 2$. This effect arises from the fact, that as soon as large scale fluctuations go non-linear, small scale fluctuations will start to grow non-linear as well and quickly regenerate an initially suppressed power spectrum on small scales. The conclusion of \cite{Boehm05A} is that the primordial power spectrum could have been exponentially surpressed initially up to mass scales of $\sim$ $10^9 M_\odot$ without being observable in large scale structure surveys today. We will now use this upper limit on the free streaming mass to find a lower limit on the particle mass or the degrees of freedom at decoupling using the prescription of eq.~(\ref{18}).
\begin{figure}[ht]
	\centering
		\includegraphics[width=8.6 cm]{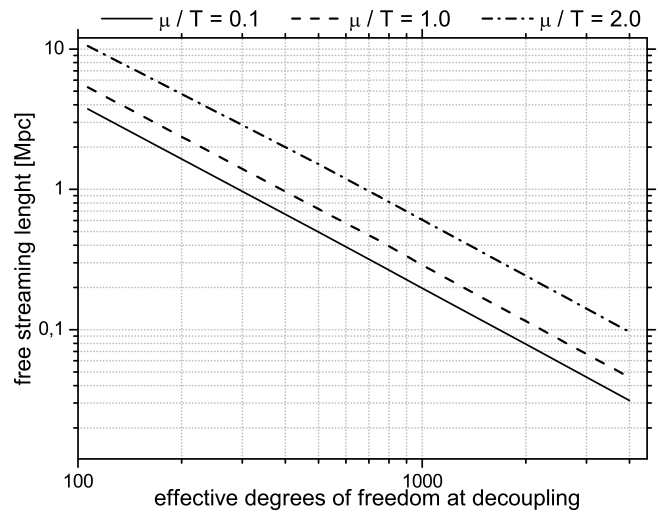}
	\caption{Free streaming length of fermions vs. the degrees of freedom at decoupling for several values of the initial chemical potential over temperature ratio}
		\label{fig:Graph10}
\end{figure}
\begin{figure}[ht]
	\centering
		\includegraphics[width=8.6 cm]{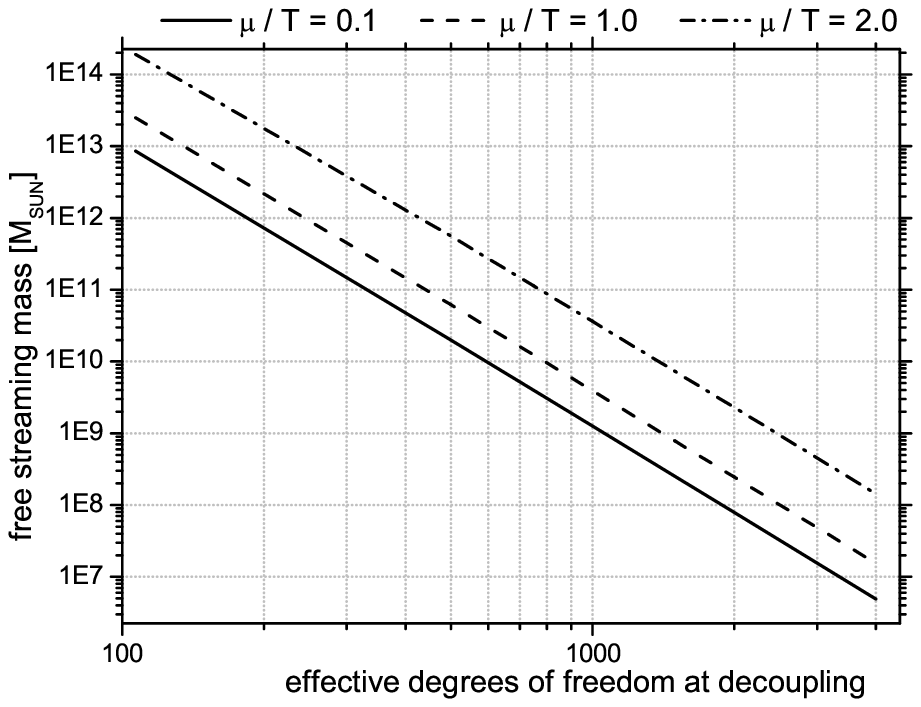}
	\caption{Free streaming mass of fermions vs. the degrees of freedom at decoupling for several values of the initial chemical potential over temperature ratio}
		\label{fig:Graph11}
\end{figure}
\begin{figure}[ht]
	\centering
		\includegraphics[width=8.6 cm]{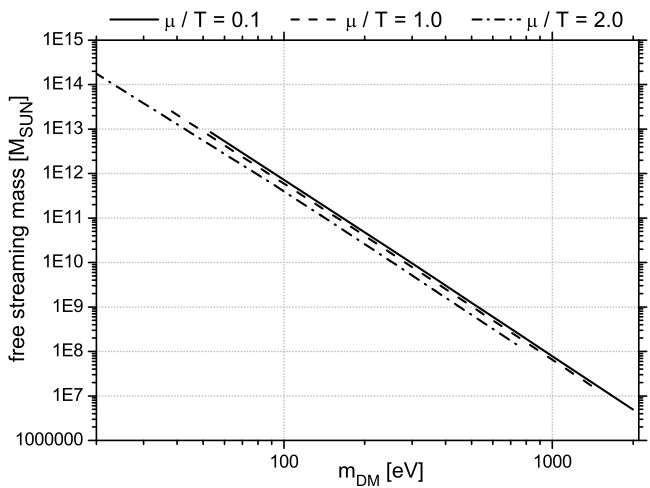}
	\caption{Free streaming mass vs. the fermion mass for several values of the initial chemical potential over temperature ratio}
		\label{fig:Graph12}
\end{figure}

In Figure 10 we show the resulting free streaming length and free streaming mass of fermions as a function of the degrees of freedom at decoupling for several values of the initial ratio of $\mu / T$. The free streaming length and mass is the same for antifermions at vanishing chemical potential and slightly lower for increasing chemical potential because antifermions will then become non-relativistic earlier as we have seen in Figure 4. Since the relative number of antifermions is also suppressed in the latter case the total damping scale should still be well described by the damping scale of fermions. One can also see that even for a $\mu / T = 2.0$ one can stay below the threshold of $\sim$ $10^9 M_\odot$ although this requires more than 2000 degrees of freedom at decoupling. Interestingly, even for a vanishing chemical potential a value of $g^s_{eff}(T_{Dec}) \approx 10^3$ (corresponding $m_{DM} \approx 500$ eV similar to the lowest mass value used in Ref.~\cite{Boehm05A}) is needed to be in agreement with Ly $\alpha$ measurements, which already exceeds the standard model value by an order of magnitude. [As a side remark: of course for a Majorana fermion $g^s_{eff}(T_{Dec})$ does only need to be half as large to reach the same mass as for a Dirac fermion with $\mu/T \approx 0$]

So structure formation gives far more restrictive bounds on the initial $\mu / T$, namely that the ratio should not exceed unity by far (keeping in mind that BBN allowed $\mu / T$ to be up to 14 for $g^s_{eff}(T_{Dec}) = 10^3$). Nevertheless one can in principle only give a lower bound on $g^s_{eff}(T_{Dec})$ since the free streaming mass is mostly fixed by the particle mass (as one can see in Figure \ref{fig:Graph12}), which increases linearly with $g^s_{eff}(T_{Dec})$, meaning that higher values of $\mu / T$ could be possible for even higher values of the degrees of freedom at decoupling ($g^s_{eff}(T_{Dec}) > 4000$). The slight decrease in the free streaming mass for rising $\mu_{DM}/T_{DM}$ at fixed $m_{DM}$ is caused by the increase in $g^s_{eff}(T_{Dec})$ in order to keep $\Omega^0_{DM}$ constant. A higher $g^s_{eff}(T_{Dec})$ means a lower temperature relative to the temperature of equilibrium particles, meaning the dark matter particles become non-relativistic at a smaller scale parameter $a_{NR}$. Remembering eqn.~(\ref{6}) this means
\begin{equation}
	a_{NR} \propto \left(\frac{T_{DM}}{T_{\gamma}}\right)_{T_{DM} > m} \propto g^s_{eff}(T_{Dec})^{-1/3}
\end{equation}
at a fixed dark matter particle mass, which explains the small reduction in $M_{FS}$.

If the dark matter distribution should not exactly coincide with the baryon distribution up to such small scales as measured by the Ly $\alpha$ data (possibly measurable by future weak lensing surveys) then this could of course relax the given restrictions.

Another way of giving a lower limit on the mass of dark matter candidates is the dark matter phase space density in galaxies. A more realistic version of the original Tremaine-Gunn-limit \cite{Tremaine79} takes into account changes of the initial particle distibutions functions during collapse to a galaxy halo as discussed by Madsen \cite{Madsen01}. He finds a lower limit on the mass of a fermionic dark matter particle of $\sim$ 380 eV, which is less restrictive than the limits from the shape of the matter power spectrum.

\subsubsection{Linear structure formation}
The simplest approach to describe the onset of structure formation after matter radiation equality is to look at the linear equation for the density contrast of matter $\delta(t,\vec{x})$ in comoving coordinates ($\vec{x} = \vec{r}/a$) which is defined by $\rho(t,\vec{x}) = \bar{\rho}(t)\left(1 + \delta(t,\vec{x})\right)$. $\bar\rho$ is the mean mass density of the corresponding matter component. In fourier space the evolution of the density contrast with time can be found by solving
\begin{equation}
	\label{20}\ddot{\delta}_k + 2H \dot{\delta}_k + \left(\frac{k^2\left\langle v^2\right\rangle}{a^2}-4\pi G\bar\rho\right)\delta_k = 0
\end{equation}
where $\delta_k(t)$ corresponds to the density contrast for one specific wave number $k$ describing density pertubations at the comoving lengthscale $\lambda_{com} = 2\pi/k$ or the physical lengthscale $\lambda_{phys} = 2\pi a/k$. For a non-relativistic collisionless fluid the velocity dispersion $\left\langle v^2\right\rangle$ replaces the square of the speed of sound $v^2_s$ of a collisional fluid and is in our approach given by (see Ref.~\cite{Hogan99} and section \ref{energy})

\begin{equation}
	\left\langle v^2\right\rangle \approx \frac{3 \bar p}{m\bar n} = \frac{9}{5} \frac{\partial \bar p}{\partial \bar\rho} \propto \left(\frac{t}{t_{eq}}\right)^{-4/3} \propto a^{-2}
\end{equation}
On the other hand
\begin{equation}
	 v^2_s = \frac{\partial \bar p}{\partial \bar\rho} = \frac{5A}{3} ~\bar\rho^{2/3} \nonumber
\end{equation}
for a collisional non-relativistic fermion fluid. One can see here that the velocity dispersion (or the speed of sound) drops like for an ordinary warm dark matter scenario (Ref.~\cite{Hogan99, Knebe03, Boehm05A, Narayanan00}) contrary to the generalized Chaplygin gas which has a late time growing speed of sound.\\
Now inserting the velocity dispersion into eq.~(\ref{20}) one finds the critical physical wavelength, the Jeans length,
\begin{equation}
	\lambda_{J} = \sqrt{\frac{\pi \langle v^2\rangle}{G\bar\rho}} \propto \left(\frac{t}{t_{eq}}\right)^{1/3} \propto a^{1/2}
\end{equation}
The corresponding Jeans mass reads:
\begin{equation}
	M_{J} = \frac{4\pi}{3}\left(\frac{\lambda_{J}}{2}\right)^3 \bar\rho = \frac{\pi^{5/2}\langle v^2\rangle^{3/2}}{6~G^{3/2}~\bar\rho^{1/2}} \propto \left(\frac{t}{t_{eq}}\right)^{-1} \propto a^{-3/2}
\end{equation}
Hence, the scalings derived here turn out to be the same as for standard WDM or CDM. In principle one has to do a full calculation of the linear power spectrum of density fluctuations in this model, but we expect results similar to warm dark matter models, although the final shape of the spectrum may be a bit more complicated due to the interplay of two (more or less) different dark matter components.\\    

\section{Conclusion}
We have shown that a relic fermionic particle family with sizeable chemical potential cannot be excluded even from the most recent BBN-calculations and large scale structure data. Using different constraints on the energy density at BBN we find that even for decoupling when $g^s_{eff}(T_{Dec})$ is equal to the maximum standard model value of 106.75 the initial ratio of $\mu / T$ could have been as large as 4.4 (when using the results of \cite{Cyburt05}), 4.5 (for the limits of \cite{Bernardis07}) or 6.3 (according to \cite{Mangano07, Barger03}). In this case the resulting mass range is at most 1.8 eV $\leq$ m $\leq$ 53 eV, where the upper bound scales linearly with $g^s_{eff}(T_{Dec})$.

\begin{figure}[ht]
	\centering
		\includegraphics[width=8.6 cm]{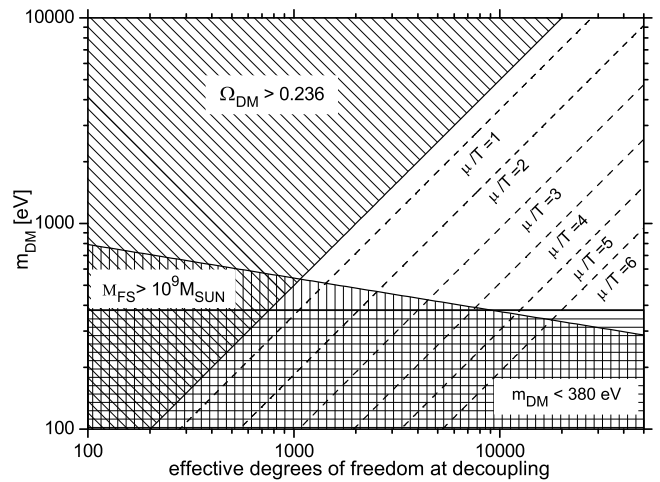}
	\caption{\itshape$g^s_{eff}(T_{Dec})$ vs. $m_{DM}$~parameter plane. Shaded regions are excluded by present day energy density (diagonal pattern), by the Tremaine-Gunn-limit (horizontal pattern) or by a too high free streaming mass (vertical pattern).\normalfont}
	\label{fig:Graph13}
\end{figure}
\noindent In figure \ref{fig:Graph13} we show the combined results on $g^s_{eff}(T_{Dec})$ and $m_{DM}$ from $\Omega_{DM}$, the free streaming mass $M_{FS}$ and the Tremaine-Gunn-limit up to very high values of $g^s_{eff}(T_{Dec}) = 5\cdot10^4$. The region with the diagonal pattern is excluded because of a too high present day energy density ($\Omega_{DM} > 0.236$ for $h = 0.7$) even for an initial $\mu/T = 0$. The border line is just the case of vanishing chemical potential (compare to eqn.~(\ref{52})) and therefore also applies to an ordinary warm dark matter relic. For all points below this line there is always one combination of $m_{DM}$, $g^s_{eff}(T_{Dec})$ and $\mu/T$ that gives the right present day energy density. The dashed lines show these right combinations for several initial values of $\mu/T$ from 1 to 6. To make this clear, a $(m_{DM}$, $g^s_{eff}(T_{Dec}))$ point on the $\mu/T = 1$ line gives the right $\Omega_{DM} = 0.236$ today for a fermion with $\mu/T = 1$ and a too high $\Omega_{DM}$ for $\mu/T > 1$ and vice versa for $\mu/T < 1$. A free streaming mass of more than $\sim10^9$ solar masses is excluded for warm dark matter as found by \cite{Boehm05A}. The region excluded by this constraint is the area with the vertical pattern. Note that the limiting line has been extrapolated beyond the $\Omega_{DM}$ limit and values of $\mu/T$ above 6. The region with the horizontal pattern is excluded by the Tremaine-Gunn-limit on the phase space density of fermions in galaxies as found by Madsen \cite{Madsen01}, which is independent of the degrees of freedom at decoupling. Fermion masses of less than 380 eV are excluded, but as one can see this limit becomes relevant only for $g^s_{eff}(T_{Dec})>10^4$. The constraints from BBN are actually so much less restrictive that they lie way outside of this plot. Remember that for $g^s_{eff}(T_{Dec}) = 106.75$ the upper bound on the initial $\mu/T$ was 4.4 (for $\Delta N_\nu < 1.63$) and for $g^s_{eff}(T_{Dec})= 10^3$ it would be even $\mu/T \leq 10.7$. Now note that the limit on the free streaming mass rules out a non-zero $\mu/T$ even for $g^s_{eff}(T_{Dec}) = 10^3$. This case would require for example only a tiny additional $\Delta N_\nu \approx 5\cdot10^{-3}$ to the number of effective neutrinos families at BBN. The BBN constraints become even less restrictive with growing $g^s_{eff}(T_{Dec})$ and thus do not appear in this summary plot. As one can see from eqn.~(\ref{55}) the maximum allowed value of $\mu/T$ grows as $(g^s_{eff}(T_{Dec}))^{1/3}$ for large values of $g^s_{eff}$.
So to summarize these results, we only find absolute lower limits on the fermionic dark matter mass and the effective degrees of freedom at dark matter decoupling. The initial ratio of chemical potential over temperature can be bounded from above for a given value of $g^s_{eff}(T_{Dec})$ but no absolute upper limit has been found.

We have seen that at least an order of magnitude more effective degrees of freedom than the standard model value are needed to allow a fermionic dark matter particle to be the dominating dark matter component (even with vanishing chemical potential) and this is also the case for an ordinary warm dark matter relic. In a gas of strings in the very early universe there are exponentially many states present (see for example Ref.~\cite{Brandenberger07,Brandenberger06}), so this would be a possible scenario to account for the very high values of $g^s_{eff}(T_{Dec})$ that are needed.

It is noteworthy that the limits on $m_{DM}$ from the free streaming mass and from the phase space limit cross, implying that the two main problems of CDM that standard warm dark matter can probably only resolve with two different WDM particle masses (i.e.~the excessive amount of small structures and the cuspy core issue in standard CDM, see e.g.~Ref.~\cite{Hogan99}) may be resolvable in the fermionic dark matter approach without this discrepancy.

Note that decoupling at $g^s_{eff}(T_{Dec}) > 100$ needs a super-weak interaction between dark matter and standard model particles, therefore constraints from cooling of supernovae and red giants that apply for light weakly interacting particles or axions do not apply here.

We have calculated the scaling of temperature, chemical potential, equation of state, energy density, number density and pressure of fermions and antifermions and have shown that even their temperatures and equations of state (as well as all other mentioned quantities) will be different in the non-relativistic regime for an initially non-zero chemical potential.
We demonstrated that the velocity dispersion, the Jeans length and the Jeans mass all have the same scaling as for regular cold or warm dark matter after adding a finite chemical potential.
  
An interesting feature about the model studied is probably that two different dark matter components can emerge from the same source, that do not necessarily clump with the same magnitude in the same region of space. This is due to a different temperature and velocity dispersion in the non-relativistic regime. A detailed numerical simulation of structure formation in such an approach would be necessary to make a clear statement about differences in the matter power spectrum today compared to a warm dark matter particle with the same mass but zero initial chemical potential. 

Finally, it is important to keep in mind that it is of course assumed that the dark matter particle was in statistical equilibrium at some point in the evolution of the universe, otherwise it will not follow a Fermi-Dirac distribution. For example in the already mentioned string gas cosmology approach such an early period of equilibrium would naturally occur. No assumption was made regarding the production mechanism and the source of the matter-antimatter asymmetry. The scope of this work was primarily to determine if a fermionic dark matter particle could in principle have a sizable chemical potential using cosmological observations and not to motivate such a relic from an underlying theory.
\\\\
\begin{acknowledgments}
We would like to thank Jens Niemeyer, Stefan Hofmann and Robert Brandenberger for constructive discussions and critical remarks. We are grateful to Paolo Salucci for bringing our attention to the cuspy-core issue. J.~Schaffner-Bielich would like to thank Richard Cyburt for conversations and insights about his BBN results. 
\end{acknowledgments}


\begin{thebibliography}{44}
\expandafter\ifx\csname natexlab\endcsname\relax\def\natexlab#1{#1}\fi
\expandafter\ifx\csname bibnamefont\endcsname\relax
  \def\bibnamefont#1{#1}\fi
\expandafter\ifx\csname bibfnamefont\endcsname\relax
  \def\bibfnamefont#1{#1}\fi
\expandafter\ifx\csname citenamefont\endcsname\relax
  \def\citenamefont#1{#1}\fi
\expandafter\ifx\csname url\endcsname\relax
  \def\url#1{\texttt{#1}}\fi
\expandafter\ifx\csname urlprefix\endcsname\relax\def\urlprefix{URL }\fi
\providecommand{\bibinfo}[2]{#2}
\providecommand{\eprint}[2][]{\url{#2}}

\bibitem[{\citenamefont{{Spergel} et~al.}(2007)\citenamefont{{Spergel}, {Bean},
  {Dor{\'e}}, {Nolta}, {Bennett}, {Dunkley}, {Hinshaw}, {Jarosik}, {Komatsu},
  {Page} et~al.}}]{Spergel07}
\bibinfo{author}{\bibfnamefont{D.~N.} \bibnamefont{{Spergel}}},
  \bibinfo{author}{\bibfnamefont{R.}~\bibnamefont{{Bean}}},
  \bibinfo{author}{\bibfnamefont{O.}~\bibnamefont{{Dor{\'e}}}},
  \bibinfo{author}{\bibfnamefont{M.~R.} \bibnamefont{{Nolta}}},
  \bibinfo{author}{\bibfnamefont{C.~L.} \bibnamefont{{Bennett}}},
  \bibinfo{author}{\bibfnamefont{J.}~\bibnamefont{{Dunkley}}},
  \bibinfo{author}{\bibfnamefont{G.}~\bibnamefont{{Hinshaw}}},
  \bibinfo{author}{\bibfnamefont{N.}~\bibnamefont{{Jarosik}}},
  \bibinfo{author}{\bibfnamefont{E.}~\bibnamefont{{Komatsu}}},
  \bibinfo{author}{\bibfnamefont{L.}~\bibnamefont{{Page}}},
  \bibnamefont{et~al.}, \bibinfo{journal}{Astrophys. J. Supp.}
  \textbf{\bibinfo{volume}{170}}, \bibinfo{pages}{377} (\bibinfo{year}{2007}),
  \eprint{astro-ph/0603449}.

\bibitem[{\citenamefont{{Tegmark} et~al.}(2006)\citenamefont{{Tegmark},
  {Eisenstein}, {Strauss}, {Weinberg}, {Blanton}, {Frieman}, {Fukugita},
  {Gunn}, {Hamilton}, {Knapp} et~al.}}]{Tegmark06}
\bibinfo{author}{\bibfnamefont{M.}~\bibnamefont{{Tegmark}}},
  \bibinfo{author}{\bibfnamefont{D.~J.} \bibnamefont{{Eisenstein}}},
  \bibinfo{author}{\bibfnamefont{M.~A.} \bibnamefont{{Strauss}}},
  \bibinfo{author}{\bibfnamefont{D.~H.} \bibnamefont{{Weinberg}}},
  \bibinfo{author}{\bibfnamefont{M.~R.} \bibnamefont{{Blanton}}},
  \bibinfo{author}{\bibfnamefont{J.~A.} \bibnamefont{{Frieman}}},
  \bibinfo{author}{\bibfnamefont{M.}~\bibnamefont{{Fukugita}}},
  \bibinfo{author}{\bibfnamefont{J.~E.} \bibnamefont{{Gunn}}},
  \bibinfo{author}{\bibfnamefont{A.~J.~S.} \bibnamefont{{Hamilton}}},
  \bibinfo{author}{\bibfnamefont{G.~R.} \bibnamefont{{Knapp}}},
  \bibnamefont{et~al.}, \bibinfo{journal}{\prd} \textbf{\bibinfo{volume}{74}},
  \bibinfo{pages}{123507} (\bibinfo{year}{2006}), \eprint{astro-ph/0608632}.

\bibitem[{\citenamefont{Steigman}(2006)}]{Steigman06}
\bibinfo{author}{\bibfnamefont{G.}~\bibnamefont{Steigman}},
  \bibinfo{journal}{Int. J. Mod. Phys.} \textbf{\bibinfo{volume}{E15}},
  \bibinfo{pages}{1} (\bibinfo{year}{2006}), \eprint{astro-ph/0511534}.

\bibitem[{\citenamefont{Astier et~al.}(2006)}]{Astier05}
\bibinfo{author}{\bibfnamefont{P.}~\bibnamefont{Astier}} \bibnamefont{et~al.}
  (\bibinfo{collaboration}{The SNLS}), \bibinfo{journal}{Astron. Astrophys.}
  \textbf{\bibinfo{volume}{447}}, \bibinfo{pages}{31} (\bibinfo{year}{2006}),
  \eprint{astro-ph/0510447}.

\bibitem[{\citenamefont{Riess et~al.}(2007)}]{Riess06}
\bibinfo{author}{\bibfnamefont{A.~G.} \bibnamefont{Riess}}
  \bibnamefont{et~al.}, \bibinfo{journal}{Astrophys. J.}
  \textbf{\bibinfo{volume}{656}} (\bibinfo{year}{2007}),
  \eprint{astro-ph/0611572}.

\bibitem[{\citenamefont{Clowe et~al.}(2006)}]{Clowe06}
\bibinfo{author}{\bibfnamefont{D.}~\bibnamefont{Clowe}} \bibnamefont{et~al.},
  \bibinfo{journal}{Astrophys. J.} \textbf{\bibinfo{volume}{648}},
  \bibinfo{pages}{L109} (\bibinfo{year}{2006}), \eprint{astro-ph/0608407}.

\bibitem[{\citenamefont{Alnes and Amarzguioui}(2007)}]{Alnes06}
\bibinfo{author}{\bibfnamefont{H.}~\bibnamefont{Alnes}} \bibnamefont{and}
  \bibinfo{author}{\bibfnamefont{M.}~\bibnamefont{Amarzguioui}},
  \bibinfo{journal}{Phys. Rev.} \textbf{\bibinfo{volume}{D75}},
  \bibinfo{pages}{023506} (\bibinfo{year}{2007}), \eprint{astro-ph/0610331}.

\bibitem[{\citenamefont{Alnes et~al.}(2006)\citenamefont{Alnes, Amarzguioui,
  and Gron}}]{Alnes05}
\bibinfo{author}{\bibfnamefont{H.}~\bibnamefont{Alnes}},
  \bibinfo{author}{\bibfnamefont{M.}~\bibnamefont{Amarzguioui}},
  \bibnamefont{and} \bibinfo{author}{\bibfnamefont{O.}~\bibnamefont{Gron}},
  \bibinfo{journal}{Phys. Rev.} \textbf{\bibinfo{volume}{D73}},
  \bibinfo{pages}{083519} (\bibinfo{year}{2006}), \eprint{astro-ph/0512006}.

\bibitem[{\citenamefont{Amendola et~al.}(2003)\citenamefont{Amendola, Finelli,
  Burigana, and Carturan}}]{Amendola03}
\bibinfo{author}{\bibfnamefont{L.}~\bibnamefont{Amendola}},
  \bibinfo{author}{\bibfnamefont{F.}~\bibnamefont{Finelli}},
  \bibinfo{author}{\bibfnamefont{C.}~\bibnamefont{Burigana}}, \bibnamefont{and}
  \bibinfo{author}{\bibfnamefont{D.}~\bibnamefont{Carturan}},
  \bibinfo{journal}{JCAP} \textbf{\bibinfo{volume}{0307}}, \bibinfo{pages}{005}
  (\bibinfo{year}{2003}), \eprint{astro-ph/0304325}.

\bibitem[{\citenamefont{{Bertolami}}(2005)}]{Bertolami05}
\bibinfo{author}{\bibfnamefont{O.}~\bibnamefont{{Bertolami}}}, in
  \emph{\bibinfo{booktitle}{ESA Special Publication}}, edited by
  \bibinfo{editor}{\bibfnamefont{F.}~\bibnamefont{{Favata}}},
  \bibinfo{editor}{\bibfnamefont{J.}~\bibnamefont{{Sanz-Forcada}}},
  \bibinfo{editor}{\bibfnamefont{A.}~\bibnamefont{{Gim{\'e}nez}}},
  \bibnamefont{and}
  \bibinfo{editor}{\bibfnamefont{B.}~\bibnamefont{{Battrick}}}
  (\bibinfo{year}{2005}), vol. \bibinfo{volume}{588} of
  \emph{\bibinfo{series}{ESA Special Publication}}, pp.
  \bibinfo{pages}{343--+}.

\bibitem[{\citenamefont{{Gorini} et~al.}(2005)\citenamefont{{Gorini},
  {Moschella}, {Kamenshchik}, and {Pasquier}}}]{Gorini05}
\bibinfo{author}{\bibfnamefont{V.}~\bibnamefont{{Gorini}}},
  \bibinfo{author}{\bibfnamefont{U.}~\bibnamefont{{Moschella}}},
  \bibinfo{author}{\bibfnamefont{A.}~\bibnamefont{{Kamenshchik}}},
  \bibnamefont{and}
  \bibinfo{author}{\bibfnamefont{V.}~\bibnamefont{{Pasquier}}}, in
  \emph{\bibinfo{booktitle}{General Relativity and Gravitational Physics}},
  edited by \bibinfo{editor}{\bibfnamefont{G.}~\bibnamefont{{Espositio}}},
  \bibinfo{editor}{\bibfnamefont{G.}~\bibnamefont{{Lambiase}}},
  \bibinfo{editor}{\bibfnamefont{G.}~\bibnamefont{{Marmo}}},
  \bibinfo{editor}{\bibfnamefont{G.}~\bibnamefont{{Scarpetta}}},
  \bibnamefont{and} \bibinfo{editor}{\bibfnamefont{G.}~\bibnamefont{{Vilasi}}}
  (\bibinfo{year}{2005}), vol. \bibinfo{volume}{751} of
  \emph{\bibinfo{series}{American Institute of Physics Conference Series}}, pp.
  \bibinfo{pages}{108--125}.

\bibitem[{\citenamefont{Sen and Scherrer}(2005)}]{Sen05}
\bibinfo{author}{\bibfnamefont{A.~A.} \bibnamefont{Sen}} \bibnamefont{and}
  \bibinfo{author}{\bibfnamefont{R.~J.} \bibnamefont{Scherrer}},
  \bibinfo{journal}{Phys. Rev.} \textbf{\bibinfo{volume}{D72}},
  \bibinfo{pages}{063511} (\bibinfo{year}{2005}), \eprint{astro-ph/0507717}.

\bibitem[{\citenamefont{Milgrom}(1983)}]{Milgrom83}
\bibinfo{author}{\bibfnamefont{M.}~\bibnamefont{Milgrom}},
  \bibinfo{journal}{Astrophys. J.} \textbf{\bibinfo{volume}{270}}
  (\bibinfo{year}{1983}).

\bibitem[{\citenamefont{Bekenstein}(2004)}]{Bekenstein04}
\bibinfo{author}{\bibfnamefont{J.~D.} \bibnamefont{Bekenstein}},
  \bibinfo{journal}{Phys. Rev.} \textbf{\bibinfo{volume}{D70}},
  \bibinfo{pages}{083509} (\bibinfo{year}{2004}), \eprint{astro-ph/0403694}.

\bibitem[{\citenamefont{{Slosar} et~al.}(2005)\citenamefont{{Slosar},
  {Melchiorri}, and {Silk}}}]{Slosar05}
\bibinfo{author}{\bibfnamefont{A.}~\bibnamefont{{Slosar}}},
  \bibinfo{author}{\bibfnamefont{A.}~\bibnamefont{{Melchiorri}}},
  \bibnamefont{and} \bibinfo{author}{\bibfnamefont{J.~I.}
  \bibnamefont{{Silk}}}, \bibinfo{journal}{Phys. Rev. D}
  \textbf{\bibinfo{volume}{72}}, \bibinfo{pages}{101301}
  (\bibinfo{year}{2005}), \eprint{astro-ph/0508048}.

\bibitem[{\citenamefont{{Cooperstock} and {Tieu}}(2006)}]{Cooperstock06}
\bibinfo{author}{\bibfnamefont{F.~I.} \bibnamefont{{Cooperstock}}}
  \bibnamefont{and} \bibinfo{author}{\bibfnamefont{S.}~\bibnamefont{{Tieu}}}
  (\bibinfo{year}{2006}), \eprint{astro-ph/0610370}.

\bibitem[{\citenamefont{{Klypin} et~al.}(1999)\citenamefont{{Klypin},
  {Kravtsov}, {Valenzuela}, and {Prada}}}]{Klypin99}
\bibinfo{author}{\bibfnamefont{A.}~\bibnamefont{{Klypin}}},
  \bibinfo{author}{\bibfnamefont{A.~V.} \bibnamefont{{Kravtsov}}},
  \bibinfo{author}{\bibfnamefont{O.}~\bibnamefont{{Valenzuela}}},
  \bibnamefont{and} \bibinfo{author}{\bibfnamefont{F.}~\bibnamefont{{Prada}}},
  \bibinfo{journal}{Astrophys. J.} \textbf{\bibinfo{volume}{522}},
  \bibinfo{pages}{82} (\bibinfo{year}{1999}), \eprint{astro-ph/9901240}.

\bibitem[{\citenamefont{{Moore} et~al.}(1999)\citenamefont{{Moore}, {Ghigna},
  {Governato}, {Lake}, {Quinn}, {Stadel}, and {Tozzi}}}]{Moore99}
\bibinfo{author}{\bibfnamefont{B.}~\bibnamefont{{Moore}}},
  \bibinfo{author}{\bibfnamefont{S.}~\bibnamefont{{Ghigna}}},
  \bibinfo{author}{\bibfnamefont{F.}~\bibnamefont{{Governato}}},
  \bibinfo{author}{\bibfnamefont{G.}~\bibnamefont{{Lake}}},
  \bibinfo{author}{\bibfnamefont{T.}~\bibnamefont{{Quinn}}},
  \bibinfo{author}{\bibfnamefont{J.}~\bibnamefont{{Stadel}}}, \bibnamefont{and}
  \bibinfo{author}{\bibfnamefont{P.}~\bibnamefont{{Tozzi}}},
  \bibinfo{journal}{Astrophys. J. L.} \textbf{\bibinfo{volume}{524}},
  \bibinfo{pages}{L19} (\bibinfo{year}{1999}), \eprint{astro-ph/9907411}.

\bibitem[{\citenamefont{{Gentile} et~al.}(2004)\citenamefont{{Gentile},
  {Salucci}, {Klein}, {Vergani}, and {Kalberla}}}]{Gentile04}
\bibinfo{author}{\bibfnamefont{G.}~\bibnamefont{{Gentile}}},
  \bibinfo{author}{\bibfnamefont{P.}~\bibnamefont{{Salucci}}},
  \bibinfo{author}{\bibfnamefont{U.}~\bibnamefont{{Klein}}},
  \bibinfo{author}{\bibfnamefont{D.}~\bibnamefont{{Vergani}}},
  \bibnamefont{and}
  \bibinfo{author}{\bibfnamefont{P.}~\bibnamefont{{Kalberla}}},
  \bibinfo{journal}{{MNRAS}} \textbf{\bibinfo{volume}{351}},
  \bibinfo{pages}{903} (\bibinfo{year}{2004}), \eprint{astro-ph/0403154}.

\bibitem[{\citenamefont{{Salucci}}(2007)}]{Salucci07}
\bibinfo{author}{\bibfnamefont{P.}~\bibnamefont{{Salucci}}}
  (\bibinfo{year}{2007}), \eprint{arXiv:0707.4370}.

\bibitem[{\citenamefont{{Gentile} et~al.}(2007)\citenamefont{{Gentile},
  {Salucci}, {Klein}, and {Granato}}}]{Gentile07}
\bibinfo{author}{\bibfnamefont{G.}~\bibnamefont{{Gentile}}},
  \bibinfo{author}{\bibfnamefont{P.}~\bibnamefont{{Salucci}}},
  \bibinfo{author}{\bibfnamefont{U.}~\bibnamefont{{Klein}}}, \bibnamefont{and}
  \bibinfo{author}{\bibfnamefont{G.~L.} \bibnamefont{{Granato}}},
  \bibinfo{journal}{\\MNRAS} \textbf{\bibinfo{volume}{375}},
  \bibinfo{pages}{199} (\bibinfo{year}{2007}), \eprint{astro-ph/0611355}.
 
\bibitem[Dav{\'e} et al.(2001)]{Dave01} Dav{\'e}, R., Spergel, 
D.~N., Steinhardt, P.~J., \& Wandelt, B.~D.\ 2001, \apj, 547, 574

\bibitem[Hennawi \& Ostriker(2002)]{Hennawi02} Hennawi, J.~F., \& 
Ostriker, J.~P.\ 2002, \apj, 572, 41

\bibitem[Bode et al.(2001)]{Bode01} Bode, P., Ostriker, J.~P., 
\& Turok, N.\ 2001, \apj, 556, 93 

\bibitem[{\citenamefont{{Sagert} et~al.}(2006)\citenamefont{{Sagert}, {Hempel},
  {Greiner}, and {Schaffner-Bielich}}}]{Sagert06}
\bibinfo{author}{\bibfnamefont{I.}~\bibnamefont{{Sagert}}},
  \bibinfo{author}{\bibfnamefont{M.}~\bibnamefont{{Hempel}}},
  \bibinfo{author}{\bibfnamefont{C.}~\bibnamefont{{Greiner}}},
  \bibnamefont{and}
  \bibinfo{author}{\bibfnamefont{J.}~\bibnamefont{{Schaffner-Bielich}}},
  \bibinfo{journal}{Eur. J. Phys.} \textbf{\bibinfo{volume}{27}},
  \bibinfo{pages}{577} (\bibinfo{year}{2006}), \eprint{astro-ph/0506417}.

\bibitem[{\citenamefont{{Cyburt} et~al.}(2005)\citenamefont{{Cyburt}, {Fields},
  {Olive}, and {Skillman}}}]{Cyburt05}
\bibinfo{author}{\bibfnamefont{R.~H.} \bibnamefont{{Cyburt}}},
  \bibinfo{author}{\bibfnamefont{B.~D.} \bibnamefont{{Fields}}},
  \bibinfo{author}{\bibfnamefont{K.~A.} \bibnamefont{{Olive}}},
  \bibnamefont{and}
  \bibinfo{author}{\bibfnamefont{E.}~\bibnamefont{{Skillman}}},
  \bibinfo{journal}{Astropart. Phys.} \textbf{\bibinfo{volume}{23}},
  \bibinfo{pages}{313} (\bibinfo{year}{2005}), \eprint{astro-ph/0408033}.

\bibitem[{\citenamefont{{Mangano} et~al.}(2007)\citenamefont{{Mangano},
  {Melchiorri}, {Mena}, {Miele}, and {Slosar}}}]{Mangano07}
\bibinfo{author}{\bibfnamefont{G.}~\bibnamefont{{Mangano}}},
  \bibinfo{author}{\bibfnamefont{A.}~\bibnamefont{{Melchiorri}}},
  \bibinfo{author}{\bibfnamefont{O.}~\bibnamefont{{Mena}}},
  \bibinfo{author}{\bibfnamefont{G.}~\bibnamefont{{Miele}}}, \bibnamefont{and}
  \bibinfo{author}{\bibfnamefont{A.}~\bibnamefont{{Slosar}}},
  \bibinfo{journal}{JCAP} \textbf{\bibinfo{volume}{3}}, \bibinfo{pages}{6}
  (\bibinfo{year}{2007}), \eprint{astro-ph/0612150}.

\bibitem[{\citenamefont{{Barger} et~al.}(2003)\citenamefont{{Barger},
  {Kneller}, {Langacker}, {Marfatia}, and {Steigman}}}]{Barger03}
\bibinfo{author}{\bibfnamefont{V.}~\bibnamefont{{Barger}}},
  \bibinfo{author}{\bibfnamefont{J.~P.} \bibnamefont{{Kneller}}},
  \bibinfo{author}{\bibfnamefont{P.}~\bibnamefont{{Langacker}}},
  \bibinfo{author}{\bibfnamefont{D.}~\bibnamefont{{Marfatia}}},
  \bibnamefont{and}
  \bibinfo{author}{\bibfnamefont{G.}~\bibnamefont{{Steigman}}},
  \bibinfo{journal}{Phys. Lett. B} \textbf{\bibinfo{volume}{569}},
  \bibinfo{pages}{123} (\bibinfo{year}{2003}), \eprint{hep-ph/0306061}.

\bibitem[{\citenamefont{{de Bernardis} et~al.}(2007)\citenamefont{{de
  Bernardis}, {Melchiorri}, {Verde}, and {Jimenez}}}]{Bernardis07}
\bibinfo{author}{\bibfnamefont{F.}~\bibnamefont{{de Bernardis}}},
  \bibinfo{author}{\bibfnamefont{A.}~\bibnamefont{{Melchiorri}}},
  \bibinfo{author}{\bibfnamefont{L.}~\bibnamefont{{Verde}}}, \bibnamefont{and}
  \bibinfo{author}{\bibfnamefont{R.}~\bibnamefont{{Jimenez}}}
  (\bibinfo{year}{2007}), \eprint{arXiv:0707.4170}.

\bibitem{Brandenberger07}
R.H.~{Brandenberger}.
\newblock {String Theory, Space-Time Non-Commutativity and Structure
  Formation}.
\newblock {\em hep-th/0703173}, March 2007.

\bibitem{Brandenberger06}
R.H.~{Brandenberger}, A.~{Nayeri}, S.~P. {Patil}, and C.~{Vafa}.
\newblock {String Gas Cosmology and Structure Formation}.
\newblock {\em hep-th/0608121}, August 2006.

\bibitem[{\citenamefont{{Kneller} and {Steigman}}(2003)}]{Kneller03}
\bibinfo{author}{\bibfnamefont{J.~P.} \bibnamefont{{Kneller}}}
  \bibnamefont{and}
  \bibinfo{author}{\bibfnamefont{G.}~\bibnamefont{{Steigman}}},
  \bibinfo{journal}{Phys. Rev. D} \textbf{\bibinfo{volume}{67}},
  \bibinfo{pages}{063501} (\bibinfo{year}{2003}), \eprint{astro-ph/0210500}.

\bibitem[{\citenamefont{{Randall} and {Sundrum}}(1999)}]{Randall99}
\bibinfo{author}{\bibfnamefont{L.}~\bibnamefont{{Randall}}} \bibnamefont{and}
  \bibinfo{author}{\bibfnamefont{R.}~\bibnamefont{{Sundrum}}},
  \bibinfo{journal}{Phys. Rev. Lett.} \textbf{\bibinfo{volume}{83}},
  \bibinfo{pages}{4690} (\bibinfo{year}{1999}), \eprint{hep-th/9906064}.

\bibitem[{\citenamefont{Chaboyer and Krauss}(2002)}]{Chaboyer02}
\bibinfo{author}{\bibfnamefont{B.}~\bibnamefont{Chaboyer}} \bibnamefont{and}
  \bibinfo{author}{\bibfnamefont{L.~M.} \bibnamefont{Krauss}},
  \bibinfo{journal}{Astrophys. J.} \textbf{\bibinfo{volume}{567}},
  \bibinfo{pages}{L45} (\bibinfo{year}{2002}), \eprint{astro-ph/0201443}.

\bibitem[{\citenamefont{{Paust} et~al.}(2007)\citenamefont{{Paust}, {Chaboyer},
  and {Sarajedini}}}]{Paust07}
\bibinfo{author}{\bibfnamefont{N.~E.~Q.} \bibnamefont{{Paust}}},
  \bibinfo{author}{\bibfnamefont{B.}~\bibnamefont{{Chaboyer}}},
  \bibnamefont{and}
  \bibinfo{author}{\bibfnamefont{A.}~\bibnamefont{{Sarajedini}}},
  \bibinfo{journal}{Astrophys. J.} \textbf{\bibinfo{volume}{133}},
  \bibinfo{pages}{2787} (\bibinfo{year}{2007}), \eprint{astro-ph/0703167}.

\bibitem[{\citenamefont{{Hansen} et~al.}(2004)\citenamefont{{Hansen}, {Richer},
  {Fahlman}, {Stetson}, {Brewer}, {Currie}, {Gibson}, {Ibata}, {Rich}, and
  {Shara}}}]{Hansen04}
\bibinfo{author}{\bibfnamefont{B.~M.~S.} \bibnamefont{{Hansen}}},
  \bibinfo{author}{\bibfnamefont{H.~B.} \bibnamefont{{Richer}}},
  \bibinfo{author}{\bibfnamefont{G.~G.} \bibnamefont{{Fahlman}}},
  \bibinfo{author}{\bibfnamefont{P.~B.} \bibnamefont{{Stetson}}},
  \bibinfo{author}{\bibfnamefont{J.}~\bibnamefont{{Brewer}}},
  \bibinfo{author}{\bibfnamefont{T.}~\bibnamefont{{Currie}}},
  \bibinfo{author}{\bibfnamefont{B.~K.} \bibnamefont{{Gibson}}},
  \bibinfo{author}{\bibfnamefont{R.}~\bibnamefont{{Ibata}}},
  \bibinfo{author}{\bibfnamefont{R.~M.} \bibnamefont{{Rich}}},
  \bibnamefont{and} \bibinfo{author}{\bibfnamefont{M.~M.}
  \bibnamefont{{Shara}}}, \bibinfo{journal}{Astrophys. Supp.}
  \textbf{\bibinfo{volume}{155}}, \bibinfo{pages}{551} (\bibinfo{year}{2004}),
  \eprint{astro-ph/0401443}.

\bibitem[{\citenamefont{{Hansen} et~al.}(2007)\citenamefont{{Hansen},
  {Anderson}, {Brewer}, {Dotter}, {Fahlman}, {Hurley}, {Kalirai}, {King},
  {Reitzel}, {Richer} et~al.}}]{Hansen07}
\bibinfo{author}{\bibfnamefont{B.~M.~S.} \bibnamefont{{Hansen}}},
  \bibinfo{author}{\bibfnamefont{J.}~\bibnamefont{{Anderson}}},
  \bibinfo{author}{\bibfnamefont{J.}~\bibnamefont{{Brewer}}},
  \bibinfo{author}{\bibfnamefont{A.}~\bibnamefont{{Dotter}}},
  \bibinfo{author}{\bibfnamefont{G.~G.} \bibnamefont{{Fahlman}}},
  \bibinfo{author}{\bibfnamefont{J.}~\bibnamefont{{Hurley}}},
  \bibinfo{author}{\bibfnamefont{J.}~\bibnamefont{{Kalirai}}},
  \bibinfo{author}{\bibfnamefont{I.}~\bibnamefont{{King}}},
  \bibinfo{author}{\bibfnamefont{D.}~\bibnamefont{{Reitzel}}},
  \bibinfo{author}{\bibfnamefont{H.~B.} \bibnamefont{{Richer}}},
  \bibnamefont{et~al.} (\bibinfo{year}{2007}), \eprint{astro-ph/0701738}.

\bibitem[{\citenamefont{{Boehm} and {Schaeffer}}(2005)}]{Boehm05}
\bibinfo{author}{\bibfnamefont{C.}~\bibnamefont{{Boehm}}} \bibnamefont{and}
  \bibinfo{author}{\bibfnamefont{R.}~\bibnamefont{{Schaeffer}}},
  \bibinfo{journal}{A\&A} \textbf{\bibinfo{volume}{438}}, \bibinfo{pages}{419}
  (\bibinfo{year}{2005}), \eprint{astro-ph/0410591}.

\bibitem[{\citenamefont{{Bond} and {Szalay}}(1983)}]{Bond83}
\bibinfo{author}{\bibfnamefont{J.~R.} \bibnamefont{{Bond}}} \bibnamefont{and}
  \bibinfo{author}{\bibfnamefont{A.~S.} \bibnamefont{{Szalay}}},
  \bibinfo{journal}{Astrophys. J.} \textbf{\bibinfo{volume}{274}},
  \bibinfo{pages}{443} (\bibinfo{year}{1983}).

\bibitem[{\citenamefont{{Narayanan} et~al.}(2000)\citenamefont{{Narayanan},
  {Spergel}, {Dav$\acute{e}$}, and {Ma}}}]{Narayanan00}
\bibinfo{author}{\bibfnamefont{V.}~\bibnamefont{{Narayanan}}},
  \bibinfo{author}{\bibfnamefont{D.}~\bibnamefont{{Spergel}}},
  \bibinfo{author}{\bibfnamefont{R.}~\bibnamefont{{Dav$\acute{e}$}}},
  \bibnamefont{and} \bibinfo{author}{\bibfnamefont{C.}~\bibnamefont{{Ma}}},
  \bibinfo{journal}{Astrophys. J.} \textbf{\bibinfo{volume}{543}},
  \bibinfo{pages}{L103} (\bibinfo{year}{2000}).

\bibitem[{\citenamefont{{Boehm} et~al.}(2005)\citenamefont{{Boehm}, {Mathis},
  {Devriendt}, and {Silk}}}]{Boehm05A}
\bibinfo{author}{\bibfnamefont{C.}~\bibnamefont{{Boehm}}},
  \bibinfo{author}{\bibfnamefont{H.}~\bibnamefont{{Mathis}}},
  \bibinfo{author}{\bibfnamefont{J.}~\bibnamefont{{Devriendt}}},
  \bibnamefont{and} \bibinfo{author}{\bibfnamefont{J.}~\bibnamefont{{Silk}}},
  \bibinfo{journal}{{MNRAS}} \textbf{\bibinfo{volume}{360}},
  \bibinfo{pages}{282} (\bibinfo{year}{2005}).

\bibitem[{\citenamefont{{Tremaine} and {Gunn}}(1979)}]{Tremaine79}
\bibinfo{author}{\bibfnamefont{S.}~\bibnamefont{{Tremaine}}} \bibnamefont{and}
  \bibinfo{author}{\bibfnamefont{J.~E.} \bibnamefont{{Gunn}}},
  \bibinfo{journal}{Phys. Rev. Lett.} \textbf{\bibinfo{volume}{42}},
  \bibinfo{pages}{407} (\bibinfo{year}{1979}).

\bibitem[{\citenamefont{{Madsen}}(2001)}]{Madsen01}
\bibinfo{author}{\bibfnamefont{J.}~\bibnamefont{{Madsen}}},
  \bibinfo{journal}{Phys. Rev. D} \textbf{\bibinfo{volume}{64}},
  \bibinfo{pages}{027301} (\bibinfo{year}{2001}), \eprint{astro-ph/0006074}.

\bibitem[{\citenamefont{{Hogan}}(1999)}]{Hogan99}
\bibinfo{author}{\bibfnamefont{C.~J.} \bibnamefont{{Hogan}}}
  (\bibinfo{year}{1999}), \eprint{astro-ph/9912549}.

\bibitem[{\citenamefont{{Knebe} et~al.}(2003)\citenamefont{{Knebe},
  {Devriendt}, {Gibson}, and {Silk}}}]{Knebe03}
\bibinfo{author}{\bibfnamefont{A.}~\bibnamefont{{Knebe}}},
  \bibinfo{author}{\bibfnamefont{J.~E.~G.} \bibnamefont{{Devriendt}}},
  \bibinfo{author}{\bibfnamefont{B.~K.} \bibnamefont{{Gibson}}},
  \bibnamefont{and} \bibinfo{author}{\bibfnamefont{J.}~\bibnamefont{{Silk}}},
  \bibinfo{journal}{{MNRAS}} \textbf{\bibinfo{volume}{345}},
  \bibinfo{pages}{1285} (\bibinfo{year}{2003}), \eprint{astro-ph/0302443}.

\end{thebibliography}
\end{document}